\documentclass[preprint]{aastex61} 

\usepackage{url}
\usepackage{natbib}
\usepackage{multirow}
\usepackage{gensymb}

\usepackage[utf8]{inputenc}  
\usepackage[T1]{fontenc} 
\usepackage{listings}
\usepackage{color}
\usepackage{xcolor}

\usepackage{rotating}

\shorttitle{Bayesian analysis of ETGs}
\shortauthors{Stalder et al.}

\begin{document}

\title{Bayesian surface photometry analysis for early-type galaxies}
\author{D. H. Stalder}
\affiliation{INPE/MCTIC, S. J. dos Campos, Brazil}
	\affiliation{NIDTEC/FP-UNA, San Lorenzo, Paraguay}
\author{Reinaldo R. de Carvalho}
\affiliation{INPE/MCTIC, S. J. dos Campos, Brazil}

\author{Martin D. Weinberg}
\affiliation{University of Masschusetts/Amherst, USA}

\author{Sandro B. Rembold}
\affiliation{UFRS, Rio Grande do Sul Brazil}

\author{Tatiana C. Moura}
\affiliation{INPE/MCTIC, S. J. dos Campos, Brazil}

\author{Reinaldo R. Rosa}
\affiliation{INPE/MCTIC, S. J. dos Campos, Brazil}
\author{Neal Katz}
\affiliation{University of Masschusetts/Amherst, USA}

\correspondingauthor{D. H. Stalder}
\email{diego.stalder@inpe.br}

\begin{abstract}
  We explore the application of Bayesian image analysis to infer the
  properties of an SDSS early-type galaxy sample including AGN.  We
  use GALPHAT \citep{GALPHAT2013} with a Bayes-factor model comparison
  to photometrically infer an AGN population and verify this using
  spectroscopic signatures.  Our combined posterior sample for the
  SDSS sample reveals distinct low and high concentration modes after
  the point-source flux is modeled.  This suggests that ETG parameters
  are intrinsically bimodal.  The bimodal signature was weak when analyzed by GALFIT \citep{GalfitPeng2002,
  	GalfitPeng2010}.  This led us to create several ensembles of
  synthetic images to investigate the bias of inferred structural
  parameters and compare with GALFIT.  GALPHAT inferences are less biased, especially
  for high-concentration profiles: GALPHAT S\'ersic index $n$, $r_{e}$
  and MAG deviate from the true values by $6\%$, $7.6\%$ and $-0.03
  \,\mathrm{mag}$, respectively, while GALFIT deviates by $15\%$,
  $22\%$ and $-0.09$\, mag, respectively.  In addition, we explore the
  reliability for the photometric detection of AGN using Bayes
  factors.  For our SDSS sample with $r_{e}\ge 7.92\,$arcsec, we
  correctly identify central point sources with
  $\mathrm{Mag_{PS}}-\mathrm{Mag_{Sersic}}\le 5$ for $n\le6$ and
  $\mathrm{Mag_{PS}}-\mathrm{Mag_{Sersic}}\le 3$ for $n>6$.  The
  magnitude range increases and classification error decreases with
  increasing resolution, suggesting that this approach will excel for
  upcoming high-resolution surveys.  Future work will extend this to
  models that test hypotheses of galaxy evolution through the cosmic
  time.
\end{abstract}


\section{Introduction}

The key to testing theories of galaxy formation and evolution is the
full use of information in galaxy image data.  Algorithmic approaches
for describing two dimensional surface photometry profiles
(e.g. SExtractor \cite{BertinEtal1996}, GIM2D \cite{GIM3D1998}, GALFIT
\cite{GalfitPeng2002, GalfitPeng2010}, 2DPHOT \cite{2DPHOT2008},
GALAPAGOS \cite{Barden2012}, PyGFit \cite{ManconeEtal2013PyGfit},
IMFIT \cite{ErwinEtal2015IMFIT}) are based on maximum likelihood
estimation (MLE), or more generally, optimizing an \emph{objective}
function that differentiates between two distributions.  This approach
has some significant limitations.  Firstly, the estimated structural
parameters are affected by random and systematic errors. For example,
pixel integration, rotation and convolution techniques used to
generate model predictions, as well as the background noise,
contamination by nearby objects, initial guesses, the form of the
objective function and the models themselves, minimization algorithms
and image sizes \citep{2007GEMS, PyMorph, 2009GUO, 2011Simard,
  2014Mendel, 2017Bernardi} may cause deviations from the correct model.
Secondly, MLE is not independent of the parameter
space coordinate system.  For example, a simple change from a linear
to logarithmic parametrization will lead to a different estimate.
Therefore inferred galaxy properties using MLE fitting tools can be
affected significantly \citep{BernadiIII2003, 2009Bernardi} by seemly
innocuous changes in the problem definition.  Thirdly, the MLE
approach cannot easily select between various models for spheroids,
bulges, discs and/or point sources given a particular galaxy image.
In other words, MLE provides no relative measure of how well the model
explains the data except in the special case of nested models.
Finally, any preexisting knowledge (e.g. published results from
related surveys) is not easily treated by the ML method.  The Bayesian
approach uses the laws of conditional probability to naturally
incorporate prior knowledge of all aspects of the scientific problem,
including both expert opinion and specifics of the model definition,
to reduce the arbitrariness that often leads to bias.  Specifically,
Bayesian methods naturally incorporate random and
systematic errors, lead to coordinate-independent estimates, and
provide a robust framework for model comparison.

In recent years, Bayesian tools have become very popular for dealing
with the drawbacks of frequentist approaches, partly due to advances
in computer hardware speeds and the implementation of sophisticated
sampling algorithms like Markov Chain Monte Carlo (MCMC) and partly
due to the overwhelming evidence that this approach works! The
astronomical community is on a \emph{wave} of testing and developing
new software tools to improve the accuracy of the inferred galaxy
structural parameters and other photometric attributes and to choose
the models that best describe their light distribution
\citep{BoucheEtal2015GalPak3D,2016Robotham}. Each implementation has
its specific advantages and weaknesses. In this work we adopt GALPHAT
(GALaxy PHotometric ATtributes); GALPHAT was the first parallelized
code available and extensively tested considering simulated galaxy
images \citep[][hereafter YMK10]{GALPHAT2013}. GALPHAT is a front-end
application of a more general and powerful tool called the Bayesian
Inference Engine\footnote{This code is publicly hosted by Bitbucket:
  \url{https://bitbucket.org/mdweinberg/BIE}.}
\citep[BIE;][]{Bie2013t}.  BIE is an application based on a parallel
MCMC algorithm that, for each parameter, gives the full posterior
distribution and likelihood marginalization.
 	
MLE packages provide a best-fit parameter value and, optionally, a
covariance matrix at the ML value.  A Bayesian inference, e.g. using
GALPHAT, provides a probability distribution of parameter values,
reflecting the improved constraints provided by the data given the
prior distribution.  Both of these describe values of the model
parameters implied by the data.  However, even in the simplest
inference, we do not truly know the underlying model family.  In more
complex inferences, a complex galaxy formation hypothesis may provide
a variety of possible model choices.

Similarly, astronomers typically use a variety of analytic light
profiles.  We often do not know which of our analytic
models best explains a particular image and want to discriminate
between possible models when different analytical forms describe
physically distinct components, such as bulges and ellipticals
(S\'ersic law), discs (exponential) or active galactic nuclei (central
point source), based on the data itself.  Bayes Law allows us to pose
an inference in the space of possible models.  This yields a relative
probability of one model over its competitors.  This posterior odds
ratio is often quoted in terms of the \emph{Bayes factor}, assuming
that all models are equally probable initially.  The Bayes factor (BF)
provide a mechanism that evaluates the evidence in favor of each
considered model rather than only testing the goodness of fit. BFs
are a natural method for model selection in the Bayesian context
\citep{Jeffreys1961, RobertRaftery, wakefield2013bayesian, Bie2013t}.
In this paper, we use GALPHAT and this Bayesian model-comparison
approach for testing the hypothesis that early-type galaxies contain a
point source typical of nuclear activity using Bayes factors.  High
resolution images obtained by the Hubble Space Telescope (HST) reveal
fine details of galaxy structure. Previous studies have shown high
correlations between nuclear activity of galaxies with galaxy
structural parameters \citep{1997Faber, HoAGN, 2002NuclearCusp,
  AGNpeng, Capetti2007, 2015HongHo, BruceEtal2016}. However,
separating the faint nucleus from the bright bulge is a difficult
task.  We will show that GALPHAT identifies weak AGN sources
photometrically in images from the SPIDER sample \citep{Spider2010}.
These images are selected to be normal early-type galaxies (ETGs)
using SDSS \(\mathrm{eClass} < 0\) and \(\mathrm{fracDev}_r>0.8\).  We
show that the power of this approach dramatically increases for PSF
scales significantly smaller than the half-light radius.  Our SDSS
images have relatively low resolution, but our results suggest that
our methods will enable higher redshift AGN identification using
upcoming high-resolution capabilities of the Thirty Meter Telescope
(TMT), the Very Large Telescope (VLT) and the James West Space
Telescope (JWST).
 
This paper extends our previous work (YMK10) by applying GALPHAT to a
large sample of SDSS early-type galaxies (ETGs) and using Bayes
factors to classify ETGs with and without nuclear unresolved sources.  The
paper is organized as follows. We begin, in Section \ref{GalphatSec}, with a
brief description of GALPHAT and its application to our target sample.
We present an automated pipeline, called PyPiGALPHAT, to analyze
galaxy structural parameters, retrieve images from a given survey,
generate configuration files, run SExtractor, and finally schedule
GALPHAT analyses and manage acquired data from an HPC cluster.  We
explore the precision and reliability of our parameter inference and
model comparisons by benchmarking on a synthetic catalog similar to
our SDSS sample in Section \ref{syntethic}.  Specifically, we
measure the bias and quantify the Bayes Factor reliability 
considering an ensemble of simulated galaxy images mimicking an SDSS
galaxy sample and compare with GALFIT results. In Section
\ref{resultSDSS} we present our findings for our target SDSS sample.
We find that our ETGs are a bimodal population and correlate one of
the modes with AGN activity using an WHAN-diagram analysis.  We
summarize our findings in Section \ref{summary}.

\section{Inferring structural parameters using GALPHAT}
\label{GalphatSec}

\subsection{Overview}

The objective function is central to optimization-based methods such
as maximum likelihood and Bayesian posterior sampling.  In both cases,
an accurate result depends on a well-posed, accurate objective
function.  For Bayesian applications, the objective function is the
true likelihood function: the probability of obtaining the observed
pixel values, \(D\) given the model \(M\) and its parameter vector,
\(\mathbf{\theta}\): \(L(D|\mathbf{\theta}) =
P(D|M,\mathbf{\theta})\).  Since most solid-state digital sensors
count electrons, the probability for each pixel is a Poisson process
whose predicted counts includes the production of electrons by photons
and all other instrumental sources.  For each pixel, the source
prediction follows from a two-dimensional integration of the flux model
convolved with the optical response function for the optics including
effects of the atmosphere.  The brute-force four-dimensional
numerical quadrature is computationally unfeasible.  Most algorithms
use a combination of interpolation and DFT-based convolution to make
this tractable.  GALPHAT uses nested interpolation tables with FFT
rotation and convolution.  We have exhaustively checked the accuracy
of this method with explicit adaptive four-dimensional quadrature.
Some of these tests will be described in Section \ref{sec:bias}.

The astronomical source is described by a magnitude or flux value, a
geometric center (X, Y), a scale or half-light radius (\(r_e\)), a sky
model, and a variety of shape parameters that parametrize the light
profile. For an ETG profile described by a \citet{Sersic1963} law,
shape parameters include the S\'ersic index (\(n\)), axis ratio
(\(q=b/a\)), and position angle (PA).  Assuming a simple flat sky
background model with one parameter (SKY), we have eight parameters in
all.  As the S\'ersic index increases, the profile increases in
concentration; e.g. \(n = 1\) is the exponential disc and \(n = 4\) is
the \citet{deVauc1948} profile.  Late-type galaxies are often
described by a bulge (S\'ersic law) and a disk (exponential law).
This adds four additional parameters: a bulge to total flux ratio, a
bulge to disk length-scale ratio, and an additional axis ratio and
position angle for the second component.  This is a total of twelve
parameters.

\begin{figure}[!ht]
  \epsscale{.60}
  \plotone{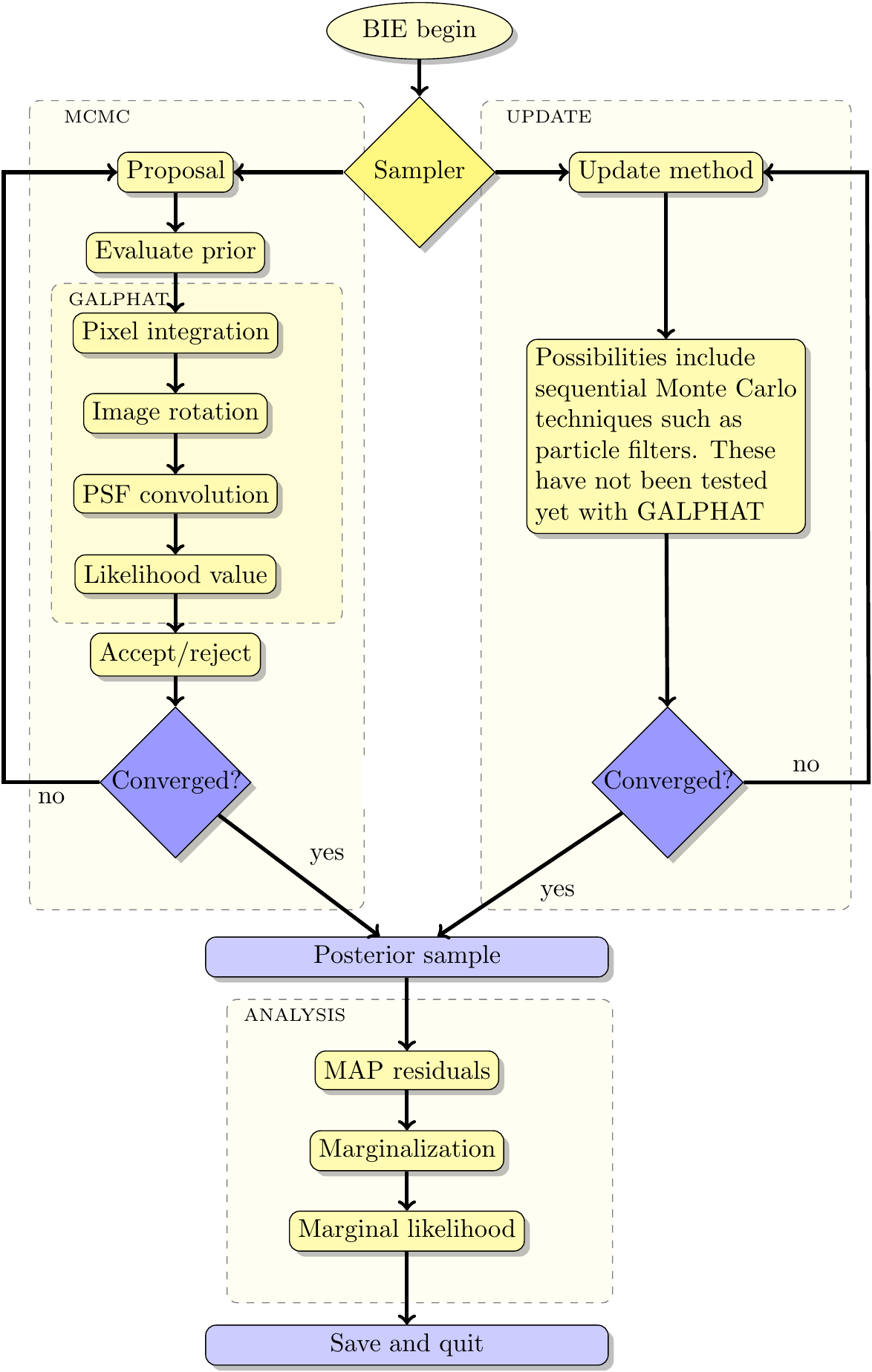}
  \caption{Graphical description of a posterior inference using BIE
    with GALPHAT.  The BIE is a general platform for estimating
    high-dimensional posterior distributions using Markov chain Monte
    Carlo (MCMC) algorithms (left column) and non-Markov chain
    samplers based on the ``update'' feature of Bayes Theorem, such as
    sequential algorithms (right column). Here, GALPHAT provides the
    likelihood-function evaluations and may be used with many methods
    provided by the BIE.  Since MCMC algorithms are the natural choice
    here, we highlight the location of the GALPHAT likelihood function
    in the computational flow.
    \label{fig:flow}}
\end{figure}

GALPHAT is part of the Bayesian Inference Engine (BIE), a general
parallel optimized software package designed 
to perform parameter inference and model
selection \citep{Bie2013t} on high-performance computing cluster
hardware.  Figure \ref{fig:flow} shows an overview of all the steps
involved in the Bayesian analysis of a galaxy image carried out in BIE
using GALPHAT.  BIE implements a variety of algorithms for sampling
and estimated the Bayesian posterior distribution.  To date these
include Markov Chain Monte Carlo (MCMC) algorithms based on the
Metropolis-Hastings concept and sequential Monte Carlo techniques
based on the Bayesian update concept, such as particle filtering.  To
date, we have only tested GALPHAT using MCMC algorithms.  For the work
described here, we will use the differential evolution algorithm
\citep{TerBraak:06}.  This algorithm has the advantage of adaptively
tuning the proposal step, removing one of the main stumbling blocks in
the application of the n\"aive Metropolis-Hastings-based method. The
BIE also includes posterior analysis for both visualization and
characterization.  The BIE implements the marginal likelihood
computation described in \citet{Weinberg:12} that we will use for
Bayes factor computation and classification in Section
\ref{recoverPS}.  The fast and accurate likelihood algorithms
implemented in GALPHAT allow one to probe the parameter space efficiently
(YWK10).  GALPHAT and the BIE produce a highly detailed description of
the relationships between all model parameters implied by the data.

However, all of this extra information comes at a price: a full
posterior simulation requires more likelihood evaluations than the
traditional optimization approach.  This application was, in part, our
motivation for developing the BIE.  Although a small number of model
inferences are tractable on workstations and laptops, a large-scale
campaign requires HPC hardware.  For this work, we developed an
automated pipeline, called PyPiGALPHAT, to streamline the work flow
for an astronomical image survey.  A detailed description of
PyPiGALPHAT is presented in Appendix \ref{pypigalphatsec}.

\begin{figure}[pht]
  \centering
  \epsscale{1.0}	
  \plotone{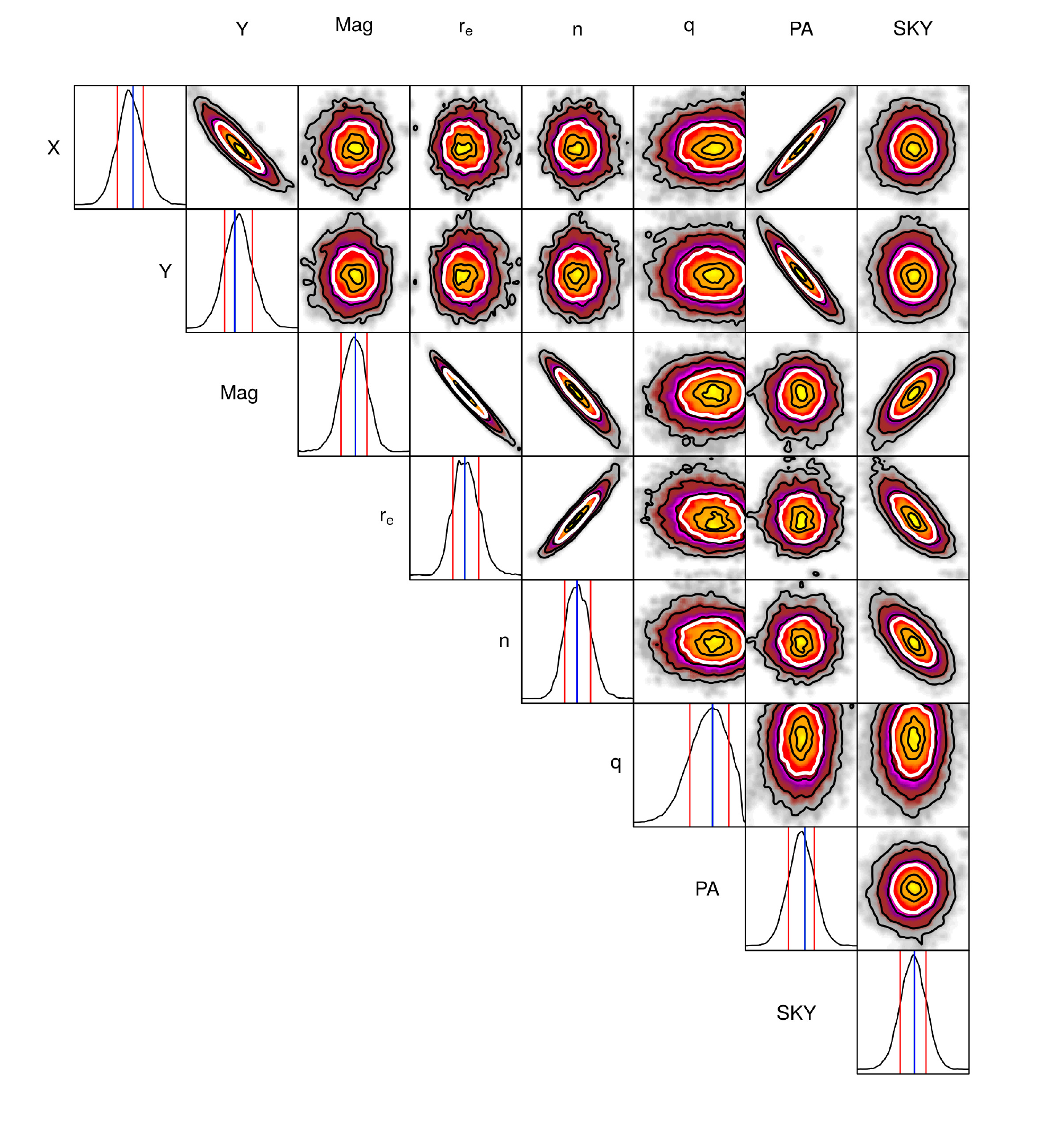}
  \caption{One- and two-dimensional marginal distributions of the
    model parameters for an example S\'ersic-model inference.  The
    diagonal contains one-dimensional marginal distributions for each
    parameter. The blue vertical line shows the position of the MAP value
    and the red vertical lines show the 1-$sigma$ range (i.e. the
    region enclosing $68.3\%$ of the probability). In the off
    diagonal, the black contours represent the 10, 30, 50, 68.3, 80,
    95 and 99$\%$ confidence levels and the white solid line is the
    68.3$\%$ confidence level.}\label{posteriorFULL}
\end{figure}

Each image analysis begins with an initial estimate of the model
parameters from the survey catalog.  The posterior distribution
describes the probability that our model's parameter vectors describe
the galaxy in light of the details provided by the observed data
including any prior analysis by the catalog compilers.  Although it is
common to characterize the inference by its best fit, the real power
in the Bayesian approach is the information provided by the full distribution
in its high-dimensional parameter space.  For example, this
distribution contains information about the full covariance of all
parameter values with no assumptions about the functional form of the
distribution.  As an example, Figure \ref{posteriorFULL} shows the
one- and two-dimensional marginal distribution of the
eight-dimensional ETG model for a S\'ersic model.  Traditional
characterizations of the posterior distribution such as the mean,
median, maximum a posteriori probability (MAP) or maximum likelihood
(ML) are straightforwardly computed by sorting the converged Markov
chain by relative probability value.  More general moments requires a
density estimate.  The BIE provides a high-dimensional kernel density
estimator based on a metric tree representation of the Markov chain
\citep[e.g.][]{Tiu.etal:2006}.

\begin{figure}[pht]
	\centering
  \includegraphics[width=0.32\textwidth]{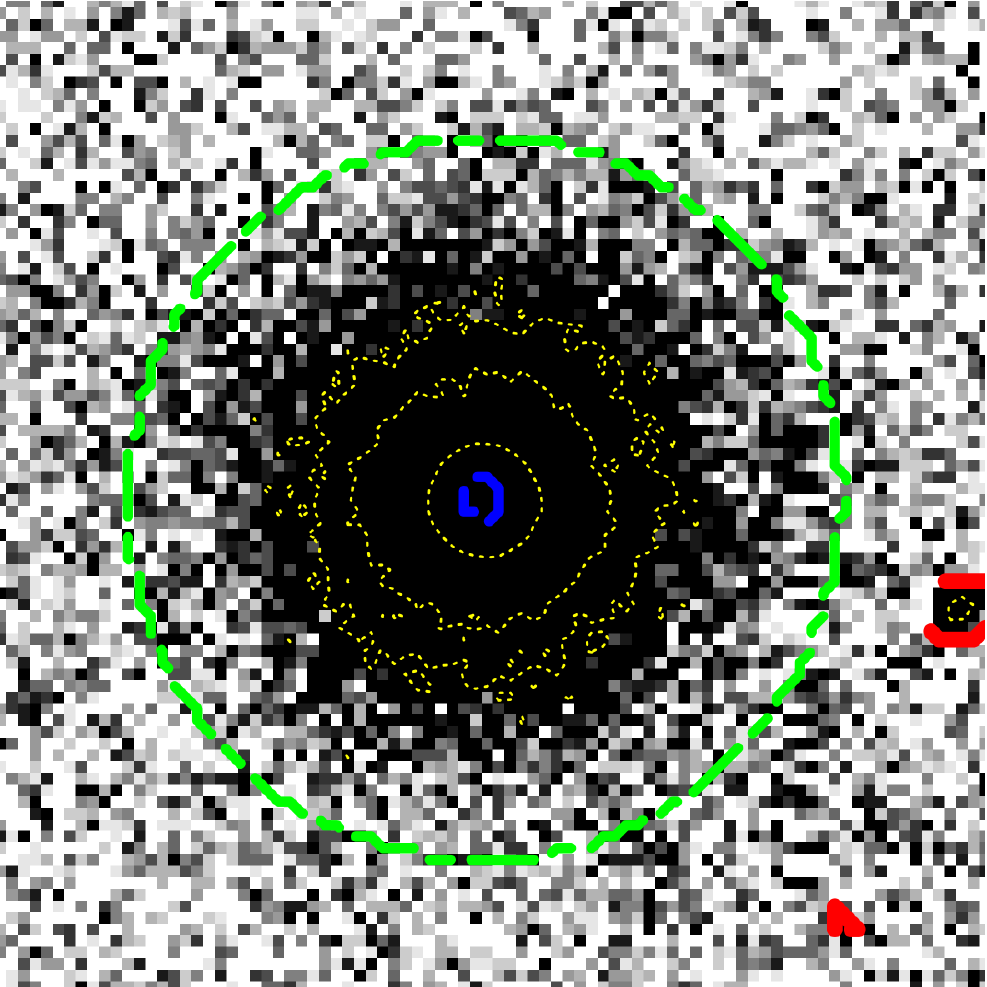} 
  \includegraphics[width=0.32\textwidth]{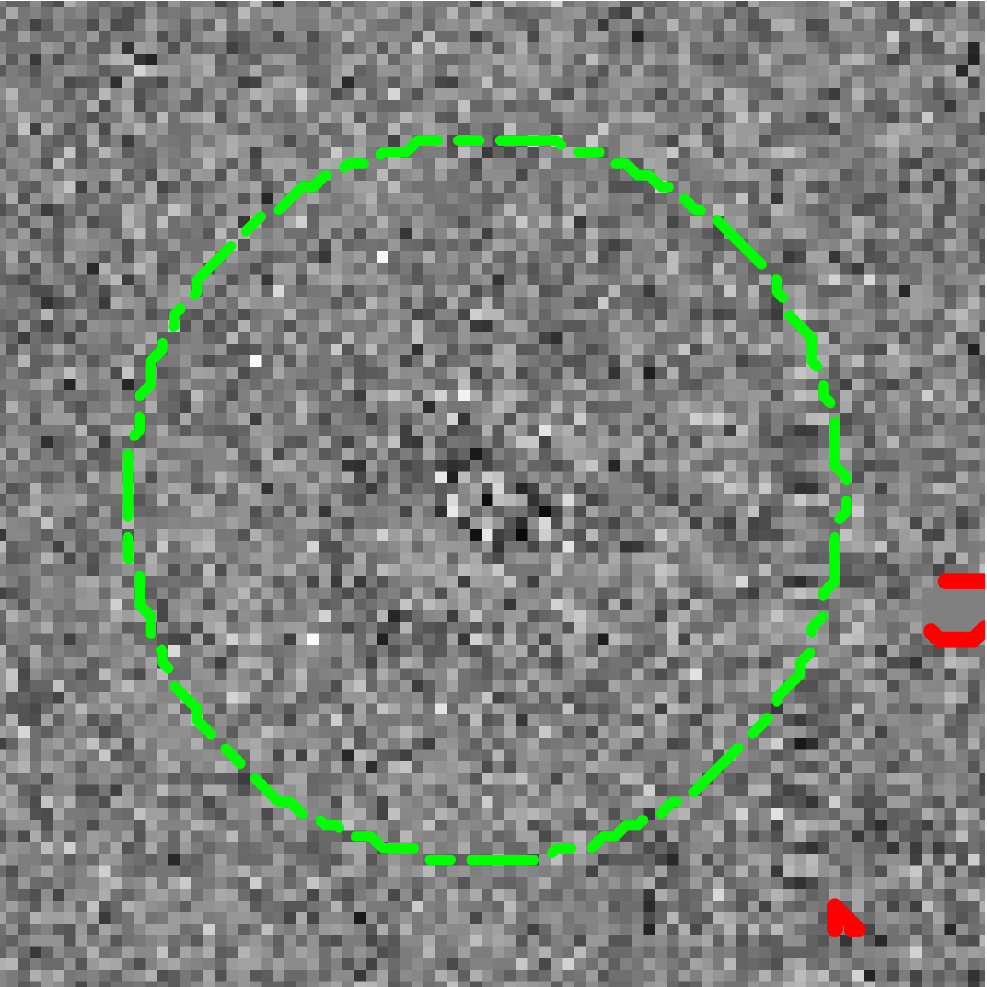} 
  \caption{SDSS postage stamp of a typical galaxy in r band with Mag =
    15.42, $r_{e}$= 6.97 \,arcsec, n = 4.53 (left) and residual images
    produced by GALPHAT with relative error varying between -2.12 and
    1.85 $\%$ (right). \label{residualcase}}
\end{figure}

Also important and often overlooked are \emph{goodness of fit}
diagnostics.  A significant goodness of fit error often indicates an
inappropriate prior assumption that may include the specified
model family itself!  The BIE includes a rigorous goodness-of-fit
based on \citet{Verdinelli.Wasserman:98}.  However, this is
prohibitively expensive for a large-scale inference campaign on an
astronomical image archive.  GALPHAT produces difference images for
the MAP and ML solutions that provide a quick visual check and a
quantitative assessment based on the chi-squared per degree of
freedom. Differences between the observed galaxy image and the model
image become obvious only when one looks at a relative residual image
(see Figure \ref{residualcase}). Large residual values and their
patterns often suggest the reason for the poor fit. Finally, if we
consider several models like S\'ersic or S\'ersic plus exponential,
GALPHAT can evaluate the evidence supporting each given model by
calculating the marginal likelihood using
the posterior distributions obtained previously.

\newpage
\subsection{Specification of the prior parameter distribution}
\label{prior}

\begin{table}[!tb]
  \begin{center}
    \caption{Prior specification.\label{stamQuality}} \vskip 0.2 truecm
    \begin{tabular}{lllllll}
      \hline\hline
      Parameters & Control & Offset & Min & Max & \multicolumn{1}{c}{Distribution} & Units \\ 
      \tableline
      X & \multirow{2}{*}{Additive} & $x_c$ & \multicolumn{1}{c}{\multirow{2}{*}{-3.0}} & \multirow{2}{*}{+3.0} & \multirow{2}{*}{Normal ($\mu=0.0$, $\sigma=1.5$)} & \multirow{2}{*}{pixels} \\  
      Y &  & $y_c$ & \multicolumn{1}{c}{} &  &  &  \\ 
      $\mathrm{Mag}$ & Additive & PetroMag &-1.0 & +1.0 & Normal ($\mu=0.0$ , $\sigma=0.2$) &  \\ 
      $r_{e}$ & Scaled & $r_{e}$ \scriptsize{(deV)} & 0.33 & 3.0 & Weibull ($k=1.21$, $\lambda=2.5$) & pixels \\ 
      $n$ & None & None & 0.5 & 14 & Normal ($\mu=6.0$, $\sigma=6.0$) &  \\ 
      $q$ & None & None & 0.09 & 0.99 & Uniform &  \\
      PA & Additive & PA \scriptsize{(deV)} & 0.0 & 0.69 & Normal ($\mu=0.0$, $\sigma=0.69$) & radians \\ 
      SKY & Scaled & SKY \scriptsize{(Sex)} & 0.97 & 1.03 & Normal ($k=1.0$, $\lambda=0.01$) & counts \\  
      \tableline
    \end{tabular}
    
    \label{priorsTable}
  \end{center}
\end{table}

Following YWK10, our prior distribution for S\'ersic model parameters
is based on a combination of linear offsets and multiplicative
scalings informed by the quoted catalog parameters and hard limits
defined by a detailed visual inspection of fitted galaxy images and
previous simulations from \citep{LaBarbera2010SpiderII}. Table
\ref{priorsTable} shows a typical set of prior distributions used for
this work. The \emph{Control} type describes whether the catalog or
input value is added to the parameter value (\emph{Additive}),
multiplied by the parameter value (\emph{Scaled}), or used directly
(\emph{None}).  The magnitude is formed from the sum of the supplied
catalog magnitude and the parameter value offset by \(-1\).  This
offset accommodates skewed prior distributions, such as the Weibull
distribution.  The offsets for $x_{c}$, $y_{c}$, $r_{e}$, PA and Mag
are set considering the reference values obtained by SExtractor and
the SDSS imaging pipeline \citep{Luptonetal2001}.

\subsection{Computational details}

\label{settingGALPHAT}

Here, we provide BIE and GALPHAT usage details specific to this
project.  Additional discussion of the BIE and GALPHAT can be found in
\cite{Bie2013t} and YMK10.  In summary, the GALPHAT likelihood
function assumes an independent per-pixel Poisson process for all
unmasked pixels.  Given a model parameter vector, GALPHAT produces a
theoretical image with the same geometry as the input
\emph{postage-stamp} FITS image and convolved with the provided PSF.
As in YMK10, we use the \emph{self-tuning} \emph{differential
  evolution} algorithm \citep{TerBraak:06} for MCMC posterior sampling
and the Gelman-Rubin \citep{Gelmarubin1992} convergence diagnostic.
Typical chain swarms have between 16 and 64 members for models with 8
to 12 parameters.  In general, fastest throughput is obtained by
assigning one chain to each core.  If physical memory is limited,
multiple chains may be assigned to each core.  We retain at least
100,000 converged chain states for each parameter estimation.
Appendix \ref{performancesec} presents performance statistics.

GALPHAT improves the computational throughput and accuracy by
maintaining a nested multi-resolution grid of the two-dimensional flux
profiles indexed by model shape parameter(s).  YMK10 tested GALPHAT
with 3000 synthetic S\'ersic galaxy images representative of the 2MASS
survey. Their ensemble contains galaxies with shape parameters varying
from 0.7 to 7, effective radius ranging from a few arc-seconds to
9.37\,arcsec (8 $\times$ the typical FWHM of the PSF) and a sky
background of $300$ [ADU].  Two multi-resolution levels provide
sufficient accuracy.  In SDSS, the scatter in the structural parameter
distributions are larger, with $n$ ranging from $2$ to $10$, and $10\%$
of all galaxies have $r_{e} \ge 10$\,arcsec \citep{Spider2010} and
hence required three multi-resolution levels.  The first grid is used to
compute the pixel flux for the inner region ($0.0396 < r_{e} <
0.396$\,arcsec), the second grid for the intermediate region ($0.396
< r_{e} < 3.96$\,arcsec), and the third is for the outer region ($3.96
< r_{e} < 39.6$\,arcsec), each one having an image size of
316.8$\times$316.8\,arcsec, 594$\times$594\,arcsec, and
792$\times$792\,arcsec, respectively. The interpolation grids are
linearly distributed as a function of $n$, ranging from 0.5 to 14.0
with 120, 120, and 240 points for the inner, intermediate, and outer
grid respectively. The flux in the central pixels ($0.0396$\,arcsec$<
r_{e}$) is computed by adaptive cubature with a strict error tolerance
in relative and absolute error of 1 part in \(10^8\).  Although
the initial grid computation is time consuming, GALPHAT automatically
caches these grids for future use.

Finally, the original GALPHAT algorithm rotated the theoretical image
to the desired position angle before convolving with the instrumental
PSF.  For very high concentration images, however, the three-shear
rotation algorithm described in YMK10 can introduce artifacts owing to
large relative flux differences between adjacent pixels. To mitigate
this problem, the PSF is inversely rotated by the desired position angle and
convolved with the non-rotated image. The image rotation is then
performed on the smoothed model image.

\section{Analysis of Simulated Images}
\label{syntethic}

\subsection{Motivation and plan}

The Bayesian posterior distributions are not guaranteed to be centered
or peaked at the true value for any single parameter or be mutually
independent of other parameters. Any preferred one-sidedness for a
statistical estimator is referred to as \emph{bias}. This definition
is only meaningful for non-Bayesian estimators, because the product of
a Bayesian inference is the posterior distribution instead of a single
parameter value Nonetheless, following astronomical tradition, we will
want to characterize the inferred parameters by some representative
value.  This might be the maximum a posterior (MAP), the median, or the
mean.  We will use the term \emph{bias} here to describe the offset of
our chosen characteristic value from the true value for an ensemble of
inferences.

The parameters for galaxy surface brightness models are often
covariate.  For example, consider the task of inferring the parameters
for a S\'ersic model for a given image.  An increase in concentration
caused by increasing the S\'ersic index \(n\) may be compensated for by
increasing the half-light radius to maintain a good fit.  Similarly,
an increase in half-light radius may be compensated for by a decrease in
magnitude.  These covariances tend to \emph{stretch} the parameter
distributions in various directions.  This stretching may be unwittingly
interpreted as a true astronomical correlation.  Using the full
posterior distribution rather than point characterization (such as
MAP, ML, mean, etc.)  prevents this source of confusion.

Finally, all galaxy image `fitting' packages produce a theoretical
image for a particular parameter vector.  Computational efficiency
demands approximation.  Approximation and truncation error in these
computed images may result in numerical artifacts that contribute to
bias and covariance.  Most of the previous studies have used the same
model generator routines to create the test images and to estimate the
structural parameters.  If one uses the same algorithm to produce the
test images and compute a comparison image for the objective function,
numerical artifacts would tend to match and indicate artificially
favorable performance.  For example, the numerical algorithms
implemented to speed up the inference process like pixel integration,
convolution, and rotation become less accurate for very concentrated
light distributions, such as those with very high S\'ersic indices ($n
\ge 8$). When the same procedure is used to generate simulated images
for testing, the errors introduced by the numerical methods will
be the same and deceptively cancel out.

To characterize the features of our inference, we updated the
independent model image generator used in YMK10 to perform the pixel
integration using explicit analytic translation and rotation
mappings and brute-force two-dimensional quadrature for the
point-spread function convolution.  This image generator implements a
recursive quad-tree cubature scheme with a strict error tolerance to
compute the pixel fluxes.  This algorithm is too expensive for GALPHAT
itself but prevents artificial concordance resulting from using the
same algorithm for image generation and image fitting.

YWK10 have shown that GALPHAT recovers the structural parameters with
a bias that depends on the image S/N, PSF FWHM, image stamp size and the
shape parameter $n$.  For comparison, we use a simplified definition
for signal-to-noise ratio that only includes noise from the astronomical source signal and the sky signal. This could be easily extended to include other
instrumental sources including read noise, dark current, etc.  We
reproduce and extend these results for testing our recent GALPHAT
improvements using the PyPiGALPHAT pipeline generator (see Appendix
\ref{pypigalphatsec}).  For future planning, we use this opportunity
to benchmark the GALPHAT performance for a modern survey (see Appendix
\ref{performancesec}).  We created various ensemble simulated images,
described in Table \ref{simulated}, to extensively test various
aspects of GALPHAT under a variety of circumstances: (1) varying
$r_{e}$ and the shape parameter $n$ as a function of S/N and PSF FWHM
(Ensemble A); (2) varying the position angle (PA) and $n$ as function
of PSF FWHM, and assuming typical values for $r_{e}$, $q$ and S/N
(Ensemble B); (3) varying the axis ratio $q$ as function of PSF FWHM,
$n$ and considering typical values for S/N and $r_{e}$ (Ensemble C);
(4) varying $n$ and PSF FWHM, for a comparison between GALPHAT and
GALFIT (Ensemble D); (5) galaxies with a real distribution of ETG
structural parameters based on the SPIDER project \cite{Spider2010}
(Ensemble E, see more details in \S\ref{recoverPHOT}); and (6)
galaxies with central point sources to test the Bayes-factor model
classification as a indicator of nuclear activity assuming typical PSF
FWHM and S/N (Ensemble F).

\begin{table}[!ht]
  \begin{center}
    \caption{Summary of simulated images ensembles.\label{simulated}}
    \vskip 0.2 truecm
    \begin{tabular}{@{} lcccr @{}}    
      \hline\hline
      Ensemble& Galaxies\tablenotemark{a}& $N_{\mathrm{realizations}}$& Parameter & Values 
      \\\tableline
      A & 1920& 2& $r_{e}$ (") & 0.99, 1.98,  2.97, 3.96, 4.95,  \\
      &  &&               &  7.92, 15.84, 31.68  \\
      & & & PSF FWHM(") & 0.75 to  2.14, steps 0.28  \\
      & & & S/N &  300, 450, 600, 750 \\
      & & & n &  2, 4, 6, 8, 10  \\
      & & & Fixed &  q$=0.7$, PA $=0$ (\degree) \\
      \tableline
      B & 360 & 2 &  PA (\degree) & -60, 0, 30, 60,\\
      & && & 90, 120, 150  \\ 
      & & & PSF FWHM(") & 0.75 to  2.14, steps 0.28   \\ 
      & & & n &  2, 4, 6, 8, 10  \\ 	
      & & & Fixed & S/N = 450, $r_{e}$=3.96 ("), \\ 	
      & & & & q = 0.7, PA = 0 (\degree) \\ \tableline
      
      C & 600 &2 &  q  & 0.5, 0.7, 0.9  \\ 
      & & & PSF FWHM(") &  0.75 to  2.14, steps 0.28    \\ 
      & & & S/N & 300, 450, 750  \\ 
      & & & n & 2, 4, 6, 8, 10   \\
      & & & $r_{e}$(") & 0.99, 3.96, 31.68  \\ 
      & & & Fixed & PA=0 (\degree) \\ \tableline
      
      D & 1200 & 50\tablenotemark{b} & n & 2, 6, 8, 10 \\ 
      & &&  PSF FWHM(") & 0.75 to  2.14, steps 0.28   \\ 
      & & & Fixed & S/N = 450, $r_{e}$=3.96 ("), \\ 	
      & & & &q = 0.7, PA = 0 (\degree) \\ \tableline
      E & 1500 & 1 & n, $r_{e}$, q, mag,FWHM & 2DPHOT Distribution (see Figure~\ref{spiderSubsample})  \\ 		\tableline
      
      F\tablenotemark{c} & 432  &1 &	$\delta \mathrm{Mag}$& 3, 5, 7, 8, 9, $\infty$\tablenotemark{d}   \\ 
      & & & $r_{e}$ (")  &  0.99, 1.98, 2.97,\\
      & & && 3.96, 7.92, 15.84 \\ 
      & &&n &4, 6, 8, 10	 \\ 
      & && q   & 0.5, 0.7, 0.9.\\ 
      & & & Fixed &  PSF FWHM = 1.3 ("),  \\ 			
      & & &      & S/N = 450, PA = 0 (\degree) \\ 			
      
      \tableline
    \end{tabular}
    \tablenotetext{a}{The total number of galaxies is $N_{r_{e}}\times N_{n}\times N_{S/N}\times N_{q}\times N_{PA}\times N_{FWHM}\times N_{\mathrm{realizations}}$.}
    
    \tablenotetext{b}{50 realizations have been generated for comparison with GALFIT.}
    \tablenotetext{c}{An ensemble to test the BF by considering the model Sésic + Point Source, where $\delta \mathrm{Mag}=\mathrm{Mag}_{PS}-\mathrm{Mag}_{sersic}$. }
    
    \tablenotetext{d}{ $\delta \mathrm{Mag}=\infty$ corresponds to a pure S\'ersic profile.}
    
  \end{center}
\end{table}

\clearpage
\subsection{Results}
\label{sec:bias}

Our simulated ensembles of galaxies allow us to measure GALPHAT biases
and uncertainties in model parameters, and the dependence of these
inferred model parameters on observational conditions such as
the point-spread function (PSF) width and the signal to noise. We combine or
\emph{pool} our posterior distributions for all galaxy inferences for
each range of parameters.  Table \ref{tab:BiasTable} presents a
summary of the differences between GALPHAT's pooled posterior median
(1-$\sigma$) and the true values as a function of $n$ and S/N. Let
\(R_{k}\) denote the value of the \(k_{th}\) percentile value for S/N.
We divide our SDSS sample into three groups as follows: small
(\(R_{0}<S/N<R_{10}\)), medium (\(R_{10}<S/N<R_{90}\)), and high
(\(R_{90}<S/N<R_{100}\)).  The median values for each S/N subgroup in
Table \ref{tab:BiasTable} were computed for galaxies with effective
radii $2.97\le r_{e} \le 4.95$\,arcsec and PSF sizes between
$1.0$\,arcsec and $1.6$\,arcsec (subsamples of the ensembles A, B and
C).  These latter two ranges are typical of the SDSS sample.  These
are low- to modest-resolution galaxy images.  Table
\ref{tab:BiasTable} shows the offset or \emph{bias} for each parameter
of the model.  We quote relative values for all quantities except for
magnitude (which is already a relative flux value) and position angle
(PA).  The relative offsets in the position of the galaxy center X, Y,
and the SKY background are below $10^{-3}$, and the axis ratio bias is
below $10^{-2}$. For most cases, the S\'ersic index, magnitude and
effective radii have relative errors smaller than $10^{-1}$. Small
relative errors indicate that the GALPHAT solution is close to the true
value.

\begin{sidewaystable}[p]
  \begin{center}
      \caption{Median and 1-$\sigma$ relative offsets in the bins
        $1.0"<\mathrm{PSF\, FWHM}<1.6"$ and considering typical SDSS
        image sizes ($2.97<r_{e}< 4.95"$)}
      \label{tab:BiasTable}
      \vspace{0.2 truecm}
      \begin{tabular}{lllllllllll}
	\hline\hline
	\multicolumn{1}{l}{}&&\textbf{$\Delta
          \mathrm{X}$}& \textbf{$\Delta
          \mathrm{Y}$} & \textbf{$\Delta
          \mathrm{Mag}$} & \textbf{$\Delta
          r_{e}/r_{e_{\mathrm{true}}}$} & \multicolumn{1}{l}{$\Delta
          n/n_{\mathrm{true}}$} & \textbf{$\Delta
          q/q_{\mathrm{true}}$} &\textbf{$\Delta \mathrm{PA}$}&
        \textbf{$\Delta \mathrm{SKY}/\mathrm{SKY}_{\mathrm{true}}$
        }\\ 
			
	S/N&$\,n$&($\times10^{-2}$) &($\times10^{-2}$) &($\times10^{-2}$) &($\times10^{-2}$) &($\times10^{-2}$) &($\times10^{-3}$) &($\times10^{-1}$) &($\times10^{-4}$)\\\tableline
			
			Low&$2$ &
		-0.14 $\pm$ 2.40 & 0.17$\pm$ 2.30 & -1.40 $\pm$ 5.0 & 2.6 $\pm$ 9.0 & 3.8 $\pm$ 6.6 & -3.5 $\pm$ 19.0 & -3.6 $\pm$ 19.0 & -1.7 $\pm$ 6.5
			\\ 
			&$4$ &
			-1.20 $\pm$ 2.00 & 0.61 $\pm$ 1.80 & -3.8 $\pm$ 4.9 & 8.2 $\pm$ 8.6 & 7.6 $\pm$ 5.9 & -11.0 $\pm$ 15.0 & -11.0 $\pm$ 19.0 & -4.9 $\pm$ 4.7
			\\ 
			&$6$&
			-0.97 $\pm$ 1.80 &0.48 $\pm$ 1.70 & -7.1 $\pm$ 6.0 & 18.0 $\pm$ 14.0 & 12.0 $\pm$ 7.4 & -13.0 $\pm$ 17.0 & -13.0 $\pm$ 23.0 & -6.5 $\pm$ 7.4\\ 
			&$8$&
		-0.96 $\pm$ 1.80 & 0.50 $\pm$ 1.80 & -7.6 $\pm$ 7.7 & 22.0 $\pm$ 22.0 & 13.0 $\pm$ 9.1 & -14.0 $\pm$ 19.0 & -19.0 $\pm$ 29.0 & -7.5 $\pm$ 14.0
			
			\\ 
			&$10$&
			-0.97 $\pm$ 1.80 & 0.49 $\pm$ 1.70 & -1.7 $\pm$ 9.1 & 7.0 $\pm$ 26.0 & 6.9 $\pm$ 10.0 & -14.0 $\pm$ 20.0 & -20.0 $\pm$ 31.0 & 2.4 $\pm$ 22.0	
			\\ 
			Med& $2$&
			-0.08 $\pm$ 1.40 & 0.03 $\pm$ 1.30 & -1.4 $\pm$ 3.6 & 2.5 $\pm$ 7.8 & 3.7 $\pm$ 5.0 & -4.0 $\pm$ 12.0 & -0.34 $\pm$ 9.2 & -2.2 $\pm$ 7.1	\\
			&$4$&
		-0.66 $\pm$ 1.2 & 0.41 $\pm$ 1.00 & -1.7 $\pm$ 2.1 & 3.6 $\pm$ 3.7 & 4.3 $\pm$ 2.8 & -5.2 $\pm$ 9.8 & -2.0 $\pm$ 8.8 & -5.2 $\pm$ 4.5
			\\ 
			&$6$ &
			-0.58 $\pm$ 1.00 & 0.41 $\pm$ 1.00 & -3.6 $\pm$ 3.0 & 8.6 $\pm$ 6.7 & 7.1 $\pm$ 3.9 & -6.8 $\pm$ 11.0 & -2.5 $\pm$ 9.8 & -7.8 $\pm$ 6.5		\\ 		
			&$8$&
			-0.54 $\pm$ 1.00 & 0.35 $\pm$ 0.95 & -5.7 $\pm$ 4.5 & 15.0 $\pm$ 12.0 & 10.0 $\pm$ 5.5 & -7.6 $\pm$ 13.0 & -2.7 $\pm$ 11.0 & -11.0 $\pm$ 12.0
			\\ 
			
			&$10$ &
	-0.52 $\pm$ 0.96 & 0.34 $\pm$ 0.94 & -4.7 $\pm$ 5.3 & 14.0 $\pm$ 16.0 & 9.1 $\pm$ 6.4 & -7.9 $\pm$ 14.0 & -3.2 $\pm$ 12.0 & -7.3 $\pm$ 16.0
			\\ 
			High& $2$ &
		-0.43 $\pm$ 1.20 & 0.24$\pm$ 0.84 & -0.15 $\pm$ 0.92 & 0.39 $\pm$ 1.0 & 1.3 $\pm$ 1.4 & -3.3 $\pm$ 4.6 & -2.3 $\pm$ 3.6 & -1.9 $\pm$ 3.4	
			\\
			
			&$4$&
	-0.26 $\pm$ 0.78 & 0.11 $\pm$ 0.59 & -1.9 $\pm$ 1.7 & 3.4 $\pm$ 2.6 & 4.6 $\pm$ 2.4 & -6.3 $\pm$ 7.1 & -1.9 $\pm$ 2.5 & -9.5 $\pm$ 7.3
			
			\\ 		
			
			&$6$&
		 -0.26 $\pm$ 0.72 & 0.09 $\pm$ 0.61 & -3.1 $\pm$ 2.3 & 6.9 $\pm$ 4.7 & 6.6 $\pm$ 3.3 & -7.6 $\pm$ 7.8 & -2.1 $\pm$ 3.0 & -11 $\pm$ 7.6
			\\ 		
			&$8$&
		-0.37 $\pm$ 0.70 & 0.19 $\pm$ 0.63 & -4.4 $\pm$ 2.6 & 12 $\pm$ 7 & 8.2 $\pm$ 3.7 & -9.3 $\pm$ 7.4 & -3.9 $\pm$ 3.8 & -15.0 $\pm$ 9.9
			\\
			
			&$10$& 
			-0.31 $\pm$ 0.71 & 0.22 $\pm$ 0.63 & -4.2 $\pm$ 2.8 & 13 $\pm$ 8.3 & 8.0 $\pm$ 3.6 & -9.8 $\pm$ 7.4 & -3.9 $\pm$ 3.8 & -10.0 $\pm$ 13.0
			\\
			\tableline		
			
		\end{tabular}
      \end{center}
  \end{sidewaystable}

Table \ref{tab:BiasTable} reveals the expected dependence of the
structural parameters on the signal-to-noise value. The relative
offset from the true value or \emph{bias} for $\mathrm{Mag}$, $r_{e}$
and $n$ decrease, albeit modestly, as S/N increases.  As $n$ increases
from 2 to 10, the bias in $\Delta \mathrm{n}$, $\Delta \mathrm{Mag}$
and $\Delta r_{e}/r_{e_{\mathrm{true}}}$ increases weakly; this is
also expected because highly concentrated (larger $n$) profiles are
more difficult to fit than profiles with lower concentration (smaller
$n$). When the measured S/N varies from low (183) to high (432), the
bias in $\Delta \mathrm{Mag}$ decreases by a factor less than 2 for
most cases. We can summarize our results as follows: (1) the centroid
is determined to order 0.0002 in all cases (e.g. 0.02\%); (2) The
magnitude is within 0.05 even for low S/N and high S\'ersic \(n\); (3)
the error in half-light radius is within 0.1 pixel in most cases; (4)
the relative error in S\'ersic \(n\) is less than 0.1 in all cases and
often much better; (5) the axis ratio error is within 0.01 in most
cases; (6) the position angle (PA) is typically within 0.2 radian
except for low S/N and high-concentration (large \(n\)) cases; (7) the
sky value is determined to better than 0.1\%.

\begin{figure}[!ht]
  \centering
  \includegraphics[width=0.99\textwidth]{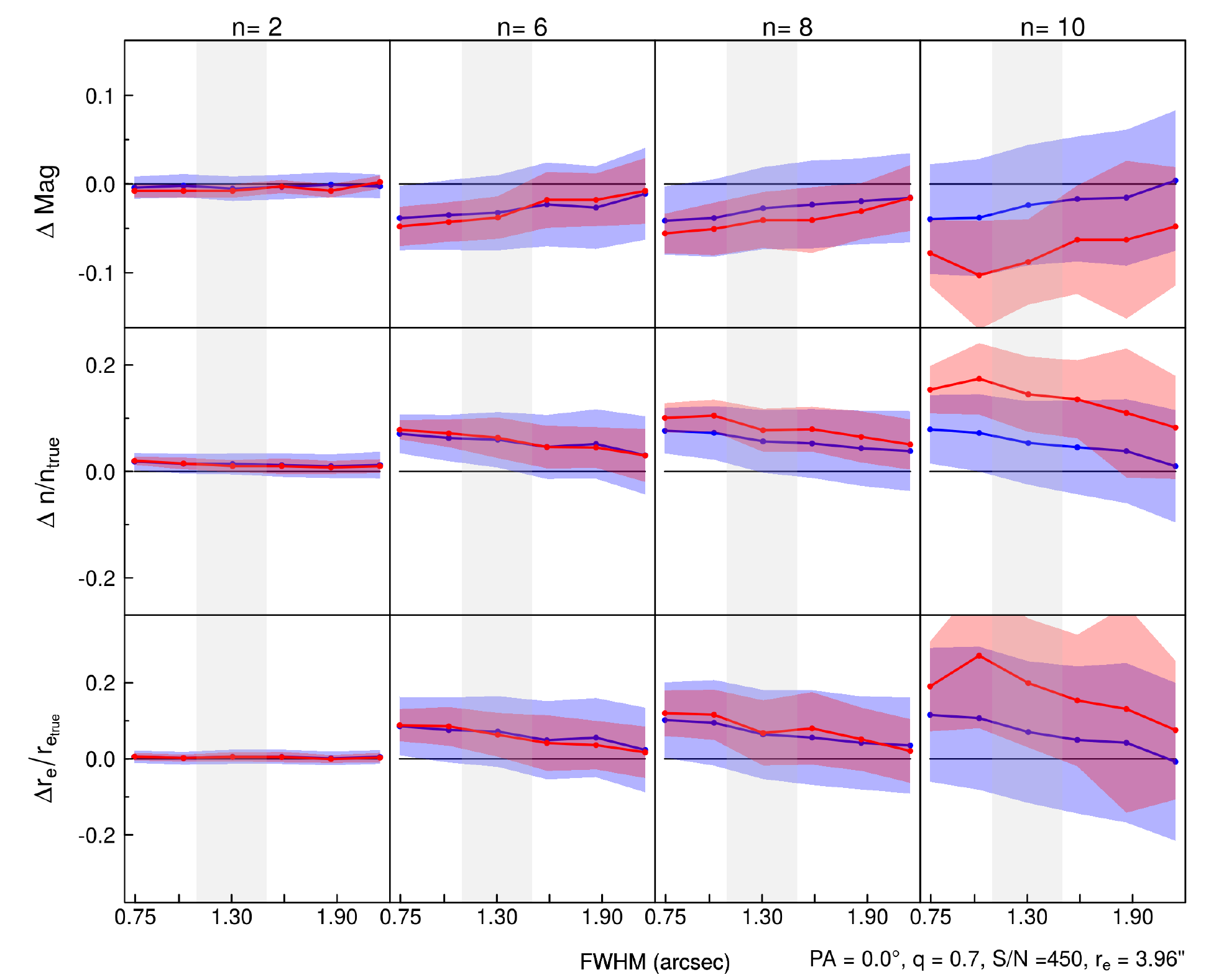}
  \caption{A comparison of relative error in GALPHAT and GALFIT for 50
    galaxy image realizations.  We show the median and the 1-$\sigma$ range
    (estimated using the interquartile range). Blue lines denote
    GALPHAT's pooled posterior medians, and red lines denote GALFIT's
    ML medians. The shaded areas indicate expected values for the PSF
    FWHM in SDSS images based on quartiles.\label{biasGalphatGalfit}}
\end{figure}

As discussed in the previous sections, GALFIT is a widely-used galaxy
image fitting tool implemented as a $\chi^{2}$-like minimization using
the Levenberg-Marquardt algorithm \citep{GalfitPeng2002,
  GalfitPeng2010}. This is a ML estimator.  In contrast, the BIE using
GALPHAT computes the probability distribution of the parameters in
light of the data given the distribution parameters before fitting.
The prior distribution includes assumptions based on physical
consistency and knowledge of the astronomical problem at hand.
Despite the philosophical and algorithmic differences between the
frequentist and Bayesian approaches, likelihood functions, and run time
performance, we can compare the accuracy of the parameter estimates
from each method.  To do this, we consider an ensemble of
simulated galaxy images that consist of 50 realizations of 24
galaxies with typical values of S/N, $r_{e}$, and $q$ (ensemble D).  We
characterize the offset from the true value using the median and
quartiles for the pooled posterior distributions for GALPHAT and using
the median and quartiles for the ML estimates from GALFIT.  This
comparison is presented in Figure \ref{biasGalphatGalfit}. For low
S\'ersic index $n$ (e.g. $n = 2$) the bias in $n$, $r_{e}$ and Mag is
negligible and both methods work similarly well. For a typical ETG
value of $n = 6$, we see a tendency of the bias being larger for
smaller values of FWHM, but we still see both methods behaving equally
well. The most striking difference appears for more extreme values of
$n$. For $n=10$, GALPHAT biases for $n$, $r_{e}$ and
$\mathrm{mag}$ are $5.7 \pm 7.9\%$, $7.6 \pm 19\%$, and $ -0.027 \pm
0.068\,\mathrm{mag}$, respectively, while GALFIT biases are $15 \pm
7.4\%$, $22 \pm 19\%$, and $-0.088 \pm 0.059\,\mathrm{mag}$,
respectively. These experiments show that GALPHAT's inference of
structural parameters is more accurate than GALFIT's, especially in the
regime of high S\'ersic index.

These results indicate that important scaling relations inferred from
structural quantities might be affected by these biases
\citep{BernadiIII2003, Shen2003, 2009Bernardi, 2017Bernardi}. For
instance, we find that the effective radius as estimated by GALFIT can
deviate from the true value by as much as 22\%. Also, the so called
Fundamental Plane (FP) of elliptical galaxies depends critically on an
unbiased evaluation of effective radius, luminosity, and velocity
dispersion. Deviations of 22\% in effective radius can certainly
compromise, for example, the study of the origin of the FP as well as
its claimed dependence on the environment.

Ideally, one uses the full posterior distribution for prediction and
inference.  However, this is not feasible for some applications and
certainly when comparing to frequentist estimators.  We now consider
whether the MAP value is sufficient to characterize the error
distribution.  Ensemble E allows us to compute the the MAP, ML,
mean and median error offset.  Each of these gives similar results for
high values of S/N.  As in our previous analysis, we combine or
\emph{pool} the GALPHAT posterior distributions for each parameter
range of interest to retain the intrinsic covariance.  Table
\ref{BiasTableNew} illustrates the error offset between the true value
and the median MAP and the median of the pooled posterior in bins of
S/N. We find that as S/N decreases, the bias increases as expected. We
see also that the pooled posterior is slightly less biased than the
median MAP value.  In short, although we expect the pooled posterior
estimates to be provide more information, the median MAP values are
adequate.

\begin{table}[ht]
  \begin{center}
    \caption{Error offsets for 1500 SDSS-like
      sample\label{BiasTableNew}} \vskip 0.2 truecm
    \scriptsize
    \begin{tabular}{ccccc}
      \hline\hline
      \multicolumn{1}{l}{Parameters}&Solution&\textbf{S/N < 183 }&\textbf{183 < S/N < 432}&\textbf{S/N > 432 }\\
      &&(Low)&(Intermediate)&(High)
      \\\tableline 
      \textbf{$\Delta \mathrm{Mag}$}& $\mathrm{Median}$ &
      -2.71 $\pm$ 3.44 &-1.66 $\pm$2.16 &-1.28 $\pm$1.09 
      \\
      &$\mathrm{Pooled\,Posterior}$ &
      -2.47	 $\pm$ 5.27 &-1.60 $\pm$ 2.75&-1.26 $\pm$ 1.39\\\tableline
			
      \textbf{$\Delta\mathrm{r_{e}}/\mathrm{r_{e_{true}}}$}&
      $\mathrm{Median}$ &
      4.50 $\pm$ 7.23 &2.86  $\pm$ 5.18  &2.64 $\pm$ 2.58 \\
      &$\mathrm{Pooled\,Posterior}$ &
      4.44 $\pm$ 11.60 &2.81 $\pm$ 6.33&2.47 $\pm$ 3.28\\\tableline
      \textbf{$\Delta \mathrm{n}$} &
      $\mathrm{Median}$ &
      3.33 $\pm$  2.73 &2.57 $\pm$ 1.95&1.74 $\pm$ 1.06\\
      &$\mathrm{Pooled\,Posterior}$ &
      3.31 $\pm$ 4.33 &2.45 $\pm$ 2.57&1.64 $\pm$ 1.21 \\\tableline
      \textbf{$\Delta \mathrm{q}/\mathrm{q_{true}}$}&
      $\mathrm{Median}$ &
      -0.22 $\pm$ 1.55 &-0.64 $\pm$ 0.95&-0.46 $\pm$ 0.66 \\
      &$\mathrm{Pooled\,Posterior}$ &
      -0.26 $\pm$ 2.23 &-0.63 $\pm$ 1.31&-0.52 $\pm$ 0.77  \\\tableline
      \tableline		
    \end{tabular}
  \end{center}
\end{table}

\begin{figure}[!ht]
  \epsscale{1.0}
  \plotone{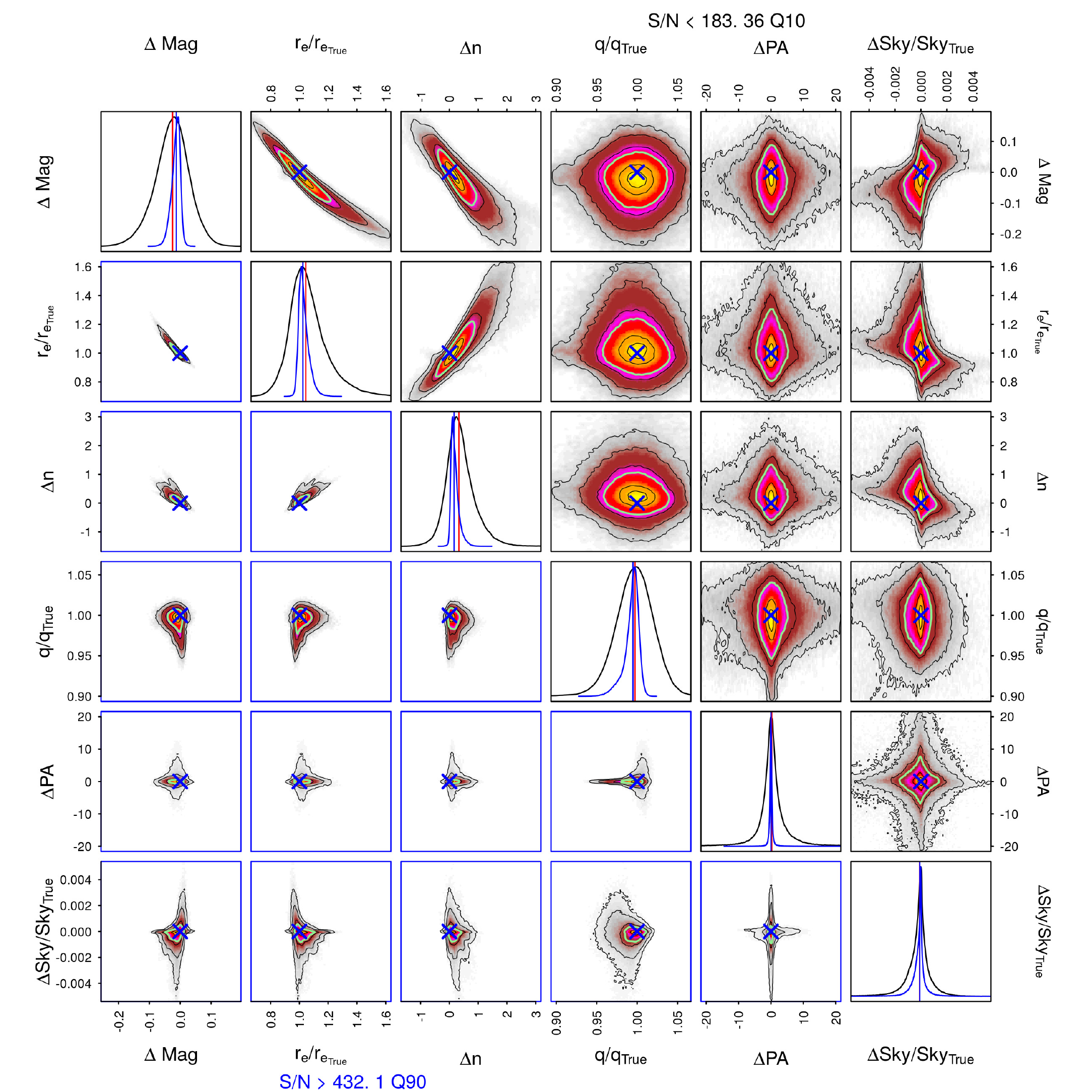}
  \caption{The posterior distribution for two subsamples of our
    simulated ensemble of ETGs for the lower 10\% in S/N (upper
    diagonal) and upper 10\% in S/N (lower diagonal).  The
    one-dimensional marginal distributions are shown on the
    diagonal for low (black) and high (blue) S/N values.  The
    parameters $r_{e}$ and q are scaled by their input values and the
    other parameters are differences from their true values. The seven
    contours represent 10, 30, 50, 80, 95, and 99$\%$ confidence
    levels. The light green solid line is the 68.3$\%$ confidence
    level. The locations of the true values are indicated by red
    vertical lines on the diagonal and cross symbols off the
    diagonal. \label{covariancesSimSNR}}
\end{figure}

To characterize the quality of GALPHAT's parameter estimates for a
realistic astronomical survey, we generated synthetic images for a
sample of 1500 ETGs with parameters randomly extracted from the 40,000
ETGs defined in the SPIDER project \citep{Spider2010}. The parameters
for each galaxy were derived using 2DPHOT \citep{2DPHOT2008}.  The
resulting sample of 1500 galaxies is Ensemble E in Table
\ref{simulated}. Figure \ref{covariancesSimSNR} illustrates the one-
and two-dimensional marginal posterior distributions for the Mag,
$r_{e}$, n, q, PA, and the SKY.  We divide the ensemble into two
subsamples containing 10\% of the total sample each: a high
signal-to-noise sample with \(S/N > 432.1\) in the lower diagonal and
a low signal-to-noise sample with \(S/N < 183.35\) in the upper
diagonal.  The diagonal contains the one-dimensional marginals 
for both subsamples coded by S/N value: black (blue) for low (high)
S/N.  In all cases, the width of the distribution is smaller for high
S/N and peaks close to the true value (shown as a red vertical line).
As expected, the half-light radius is covariant with the concentration
and magnitude: larger \(r_e\) and/or smaller S\'ersic index \(n\) may
be compensated for by a smaller MAG.  The overall trends are similar with
the analysis presented in YMK10 (Figures 5-8).

\begin{figure}[!ht]
  \epsscale{1.0}
  \plotone{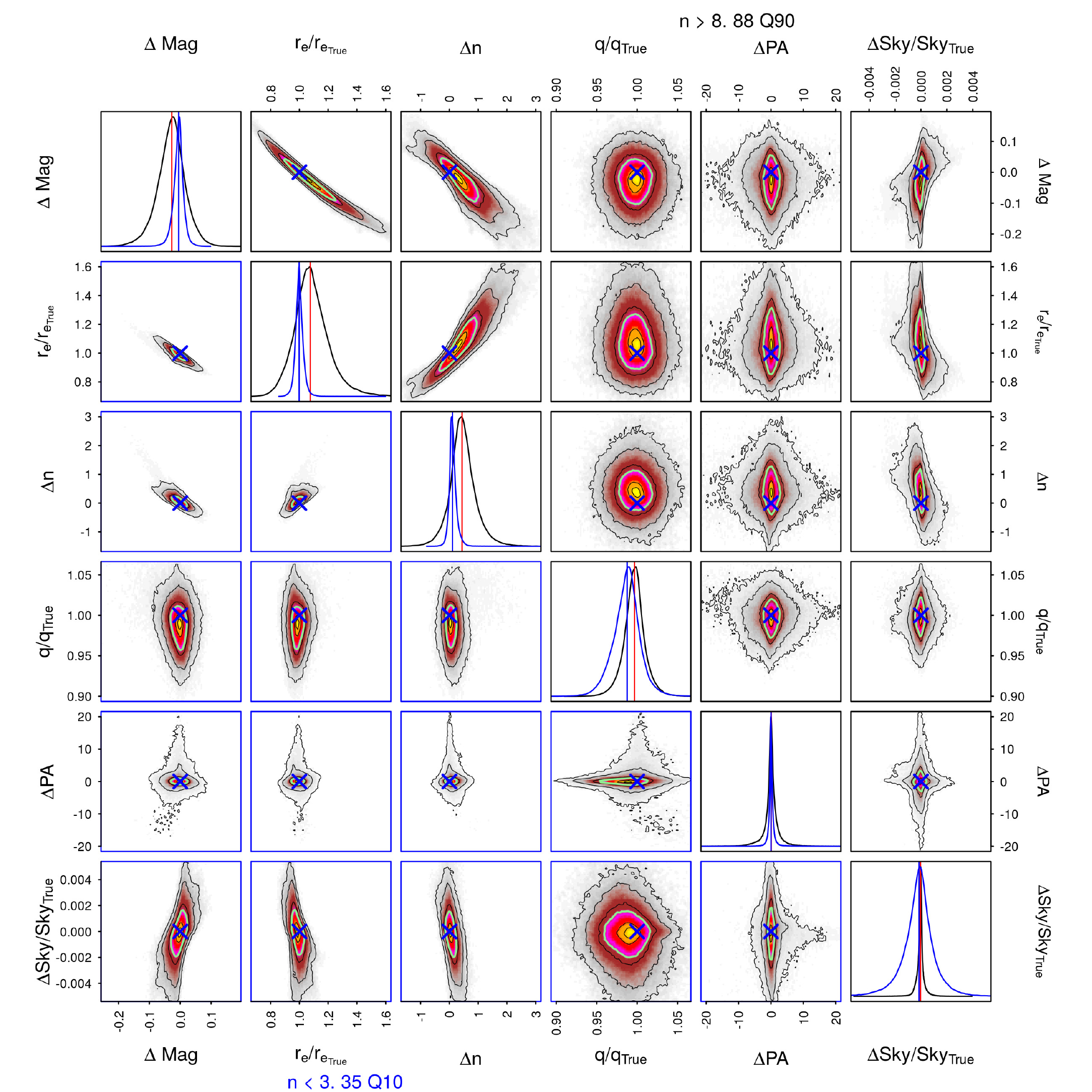}
  \caption{As in Fig. \ref{covariancesSimSNR} with the lower
    and upper 10\% in S\'ersic index \(n\).
    \label{covariancesSimN}}
\end{figure}

\begin{figure}[!ht]
  \epsscale{1.0} \plotone{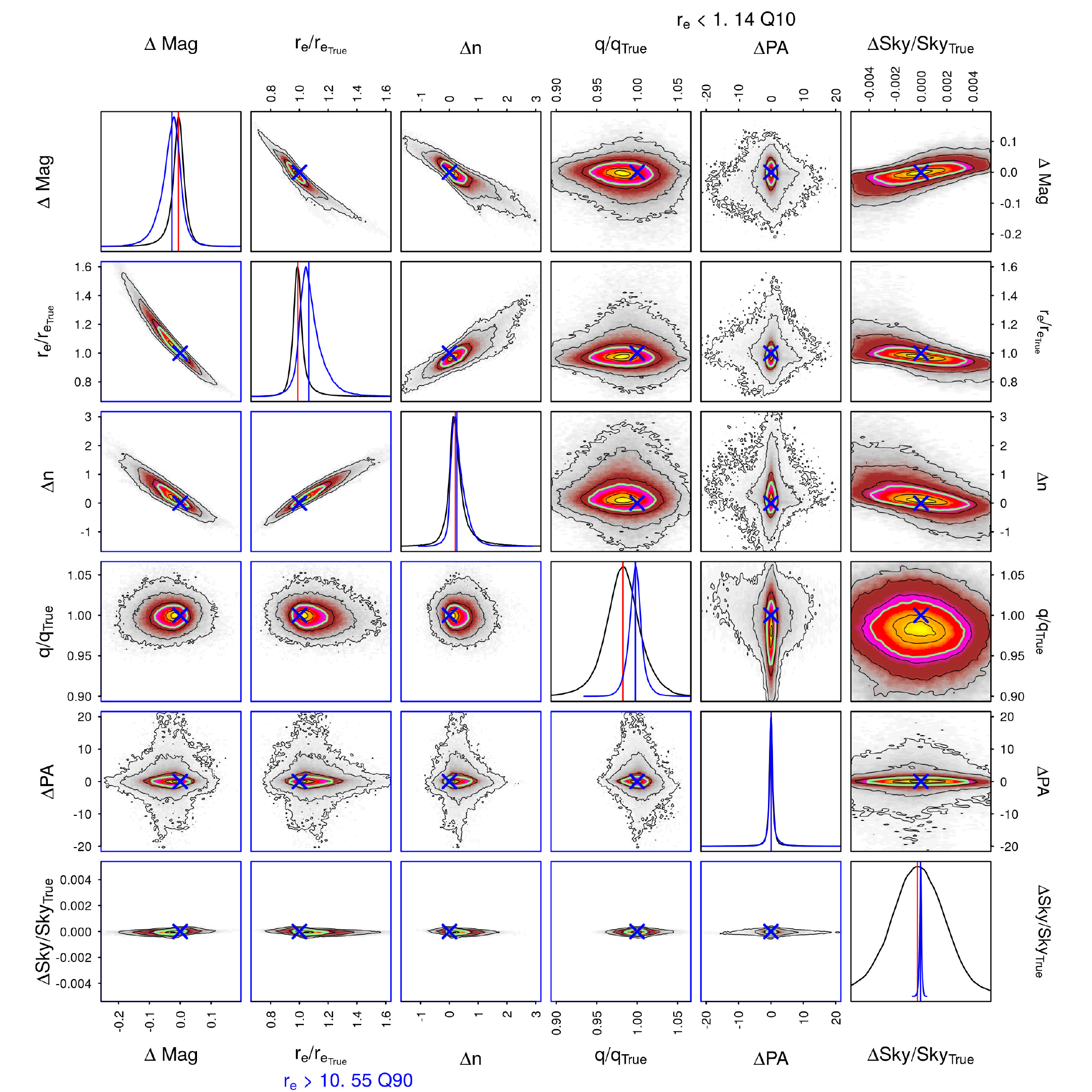}
  \caption{As in Fig. \ref{covariancesSimSNR} with the lower and
    upper 10\% in half-light radius \(r_e\).
    \label{covariancesSimRe}}
\end{figure}

\begin{figure}[!ht]
	\epsscale{1.0} \plotone{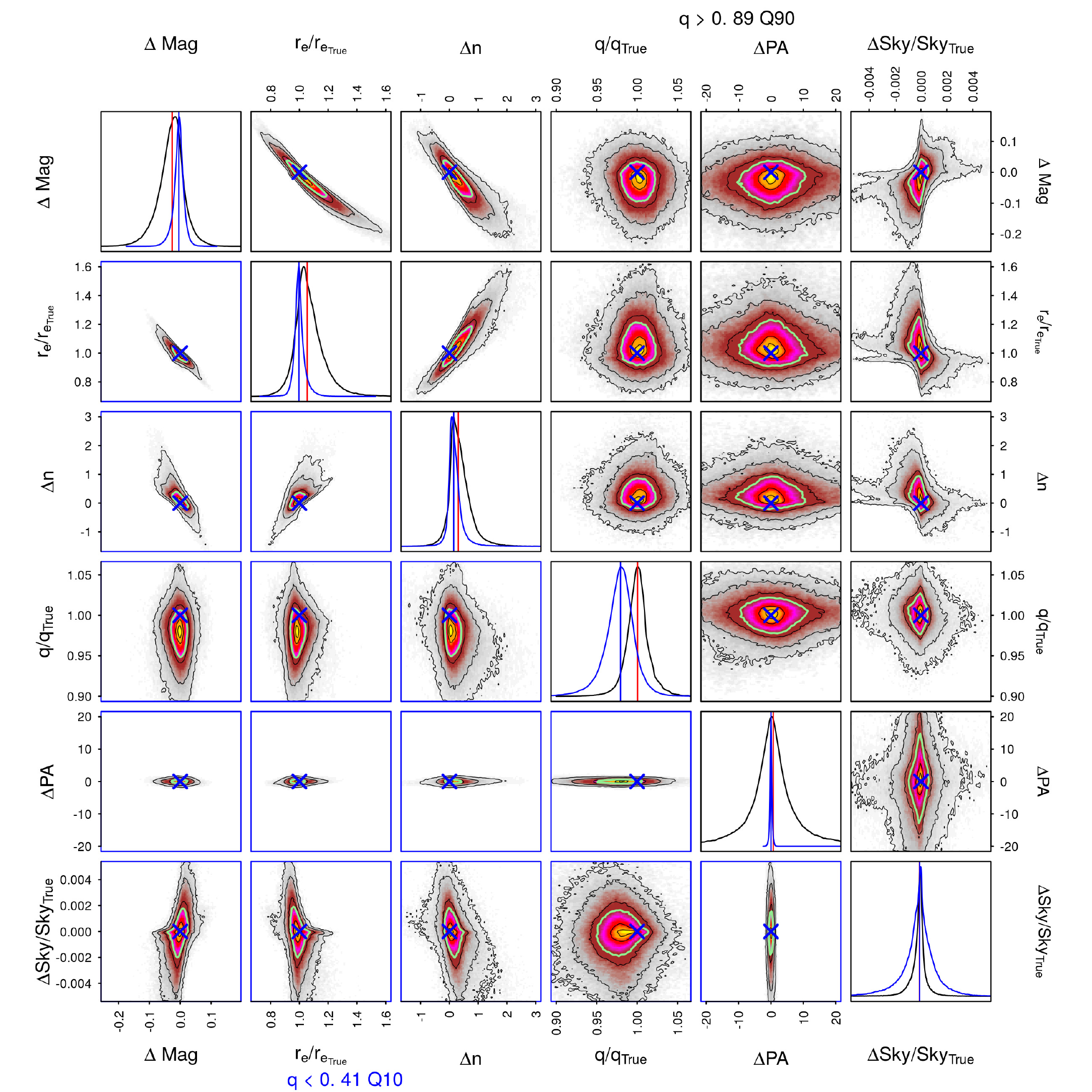}
	\caption{As in Fig. \ref{covariancesSimSNR} with the lower
          and upper 10\% on axis ratio \(q\).
          \label{covariancesSimQ}}
\end{figure}

We now select two subsamples, each containing 10\% of the total, as in
Figure \ref{covariancesSimSNR}, but selecting on extreme values of
S\'ersic index; the low-value sample has \(n<3.35\) (lower diagonal,
blue) and high-value sample has \(n>8.88\) (upper diagonal, black).
Figure \ref{covariancesSimN} reveals larger credible intervals for
high \(n\) values with similar trends in covariance as described for
Figure \ref{covariancesSimSNR}.  Figure \ref{covariancesSimRe}
describes posterior distribution for the lower 10\% and upper 10\% in
half-light radius, $r_{e}$. For small values of \(r_e\), the wings of
the galaxy profile are in the sky and smeared out by the PSF.  This
decreases the quality of the inferred value of axis ratio \(q\) and
S\'ersic index \(n\).  These findings are consistent with the previous
analysis in YMK10.  Figure \ref{covariancesSimQ} selects the sample on
the lower and upper 10\% of the axis ratio distribution.  As expected,
rounder galaxies, larger \(q\), have more poorly determined values of
position angle PA.  The low \(q\) sample is biased toward lower \(q\)
values at the 5\% level.  Conversely, the bias in $n$ and $r_{e}$ is
lower for lower axis ratios.

\clearpage

\subsection{Recovery of structural parameters for ETGs}
\label{recoverPHOT}

\begin{figure}[ht]
	\epsscale{1.1}
	\plotone{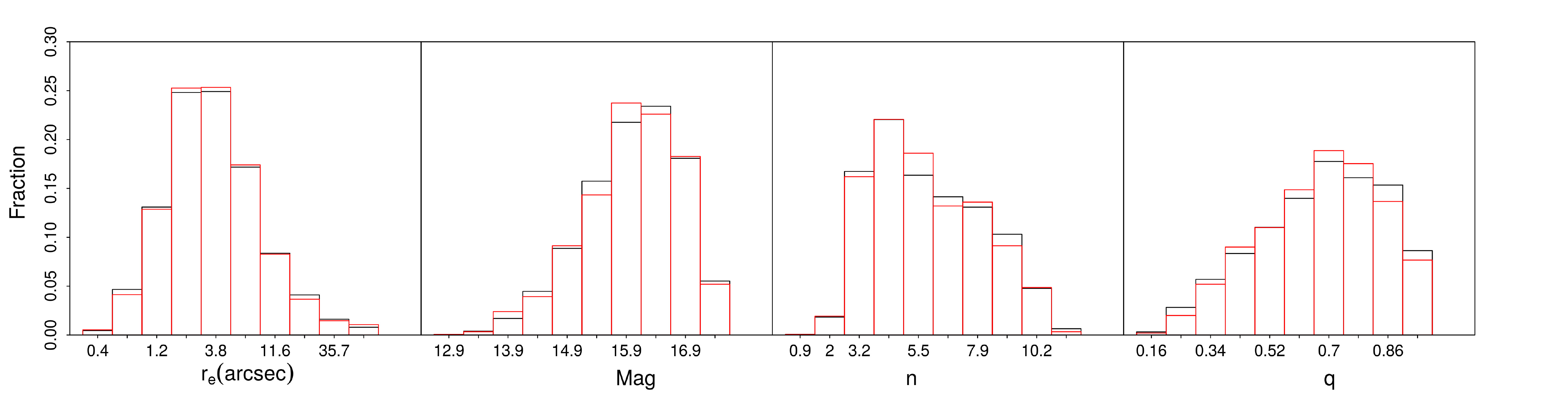}
	\caption{The distribution of structural parameters found in
		\cite{Spider2010}. Black: the distribution of parameters using
		2DPHOT \citep{2DPHOT2008} for the 40000 ETGs.  Red: the
		distribution of parameters from GALPHAT for 1500 member random
		subsample.\label{spiderSubsample}}
\end{figure}

Most astronomical hypotheses are tested using ensemble distributions.
Here, we test the recovery of the distribution properties.  We
simulate this by comparing the marginal distributions of structural
parameters (Mag, $r_{e}$, and n) inferred by GALPHAT with the original
input distribution.  Figure \ref{spiderSubsample} illustrates that our
subsample of 1500 ETGs have a distribution similar to the SPIDER sample.

For each marginal distribution from the pooled posterior, we compute the Anderson-Darling (AD), and the permutation test p-values. 
To compare the estimated marginal distributions with the true value 
distributions, we take 1000 random subsamples from the pooled posterior 
(1 point from each posterior).
Then we apply several statistical tests assuming the null hypothesis that the 
two samples have identical probability distributions. A permutation test
builds a sampling distribution for a statistic (e.g. the mean) by
randomly resampling the data and compiling a frequency distribution.
If the statistic for the test data set is in the tails of the frequency
distribution, we reject the null hypothesis of the same parent
distribution.  For our experiments, we consider 2000 samples without
replacement using the \emph{permTS} package from the R Project
\citep{FayShaw2010}. 

\begin{table}[ht]
	\begin{center}
		\caption{Distribution tests p-values for ETG sample\label{Permu}} \vskip
		0.2 truecm 
		\scriptsize
		\begin{tabular}{ccccccc}
			\hline\hline
			\multicolumn{1}{l}{Parameters}&\multicolumn{3}{c}{\textbf{Permutation}}&\multicolumn{3}{c}{\textbf{AD}}\\
			&$n_{true}$<6&$n_{true}$<8&$n_{true}$<10  &$n_{true}$<6&$n_{true}$<8&$n_{true}$<10  \\ 
			\tableline  
			\textbf{$ \mathrm{Mag}$} &
			0.61$\pm$0.25 &
			0.60$\pm$0.25      &
			0.57 $\pm$0.26   &
			0.75$\pm$0.24&
			0.75$\pm$0.24&
			0.71 $\pm$ 0.27 \\\tableline
			\textbf{$ r_{e}$} &
			0.61$\pm$0.25				&
			0.60$\pm$0.26  &
			0.50 $\pm$ 0.27&
			0.69$\pm$0.27 & 
			0.73$\pm$0.25  &
			0.64$\pm$ 0.28\\\tableline
			\textbf{q} &
			0.24$\pm$0.24 &
			0.27$\pm$0.26&
			0.37$\pm$0.28  &
			0.36$\pm$0.30&
			0.44$\pm$0.29&
			0.47$\pm$0.25 \\\tableline
			\textbf{n} &
			0.32$\pm$0.26 &
			0.28$\pm$0.26  	 &
			0.07 $\pm$ 0.11  	 &
			0.34$\pm$0.27  &
			0.23$\pm$0.24&
			0.07 $\pm$  0.12\\\tableline
		\end{tabular}
	\end{center}
\end{table}

Table \ref{Permu} presents the mean and standard deviation of p-values
for the Anderson-Darling (AD) and the permutation tests for all the pooled
posterior subsamples for a sequence of cuts in S\'ersic $n$.  Both
tests indicate that our pooled posterior subsamples are similar to
the true distributions of Mag, $r_{e}$ and $q$.  The p-values for the
S\'ersic index distribution, however, are lower, reflecting that only
a small number of subsamples (368/355 out of 1000 for the
AD/Permutation tests) do not reject the null hypothesis. This is
expected because highly concentrated (larger $n$) profiles have
significant bias. When we decrease the upper threshold from $n_{true}<10$
to $n_{true}<8$ the p-value means in the last rows are at least 3
times larger; 774 (863) out of 1000 subsamples now pass the AD
(Permutation) test. For $n_{true}<6$, the mean p-values become even
larger and more cases pass both tests.  Overall, these
statistical tests tell us that GALPHAT pooled posterior distributions
recover the true value distributions of our synthetic sample.

\subsection{Inferring the magnitude of central point sources}
\label{recoverPS}

GALPHAT can be easily extended to model ETGs with a central AGN source as a
combination of an extended galaxy light profile and a central point
source.  The computational challenge is two-fold: (1) a reliable
recovery of the point-source magnitude when it exists; and (2) a model
comparison test to discriminate galaxies with and without point
sources.  To test both the reliability of parameter estimation and
Bayesian model selection, we generate simulated galaxy images
consisting of both a S\'ersic profile and a nuclear point source (PS).
We assume that this additional component coincides with the galaxy
center and is defined by the magnitude of the central point source
only, $\mathrm{Mag}_{PS}$.

We proceed as in the previous sections: we sample the posterior
distribution using BIE and GALPHAT for a model with a point source and
a model with no point source.  After convergence, the marginal
likelihood is computed by the BIE following the algorithm given in
\citet{Weinberg:12} and the extensions discussed in
\citet{Weinberg.etal:2013}.  Given a space of models, say \(\{M_{1},
M_{2}\}\), we can use Bayes theorem to compute the relative
probability of the model given the data: \(P(M_{j}|D) = P(M_{j})
P(D|M_{j})/\sum_k P(M_{k}) P(D|M_{k})\).  The relative probability of
two models given the same data are then:
\[
BF_{12} \equiv \frac{P(M_{1}|D)}{P(M_{2}|D)} =
\left(\frac{P(M_{1})}{P(M_{2})}\right)
\left(\frac{P(D|M_{1})}{P(D|M_{2})}\right).
\]
The first term is the prior odds ratio, which we typically assume to
be unity, and the second term is called the \emph{Bayes factor}
(BF). The numerator and denominator in the Bayes factor is exactly the
likelihood function marginalized over the prior probability.  It is
often said that the BF is the relative evidence for two competing
models in the data. The evidence in the data favors one hypothesis,
relative to another, exactly to the degree that the hypothesis
predicts the observed data better than the other.  See
\citet{RobertRaftery} for more details.

As in the previous section, we will measure the reliability of
selecting a S\'ersic profile ($M_{1}$) and a S\'ersic profile and a
point source ($M_{2}$) using an ensemble of synthetic images.  The PSF
FWHM and the pixel scale are fixed by the instrumental and
observational conditions of our SDSS sample.  Our test ensemble of 432
galaxies covers a wide range of structural parameters designed to
assess the limitations of the BF analysis for identifying a nuclear
point source.  Of these, 360 of 432 cases have a central point source
with varying relative magnitudes, \(\delta\mathrm{Mag}\), and 72 of 432
cases have a single S\'ersic profile.  For each combination of
effective radius \(r_e\) and S\'ersic index \(n\), we choose discrete
values \(\delta\mathrm{Mag}_{true}= \mathrm{Mag}_{PS_{true}}-\delta
\mathrm{Mag}_{sersic_{true}}=\{3,5,7,8,9\}\).

\begin{figure}[!ht]
  \centering
  \epsscale{1.0}
  \plotone{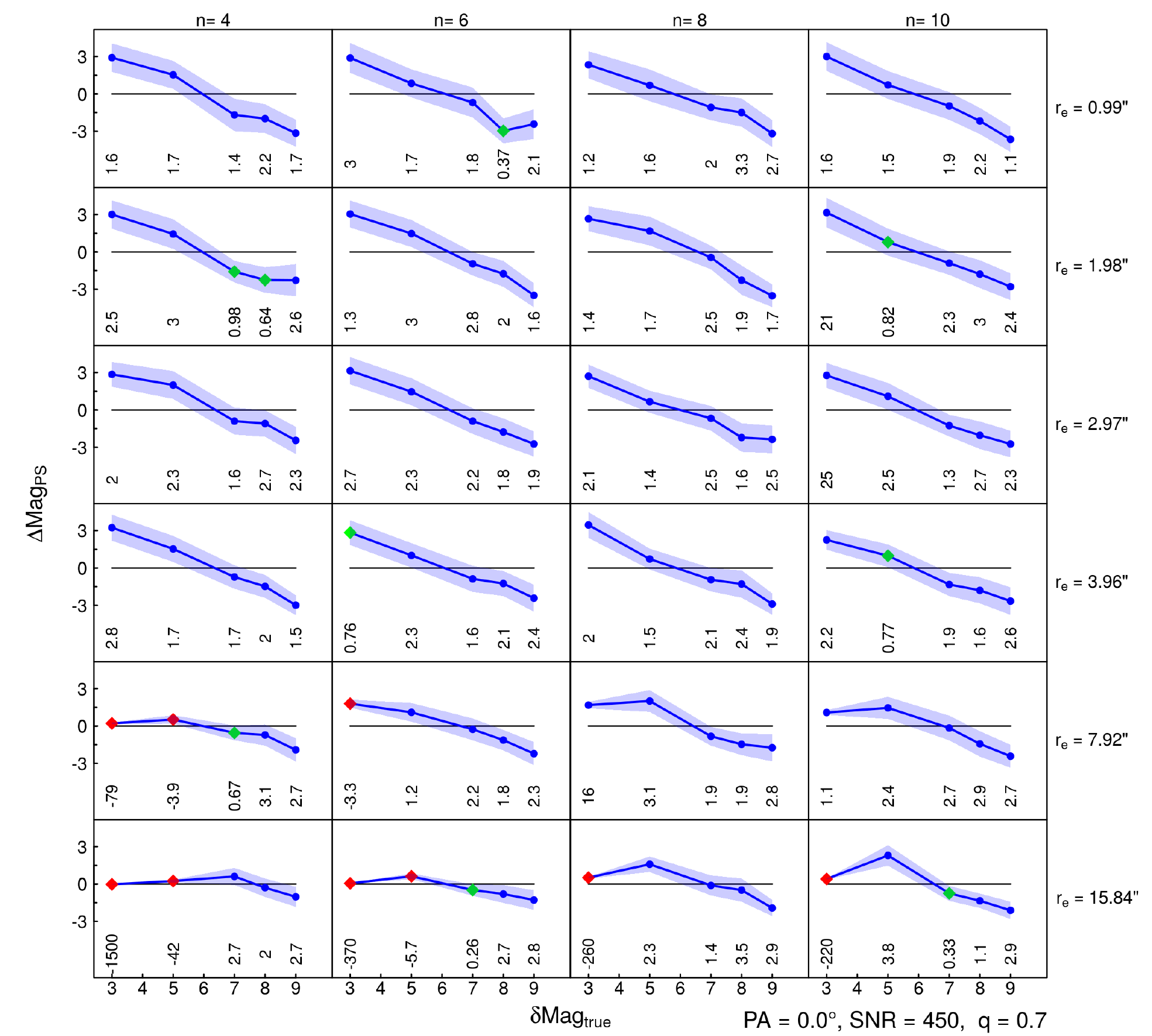}
  \caption{Bias in inferred point-source magnitude,
    \(\Delta\mathrm{Mag}_{\mathrm{PS}}\), as function of the true
    magnitude difference, \(\delta \mathrm{Mag}_{\mathrm{true}} =
    \mathrm{Mag}_{PS_{\mathrm{true}}} -
    \mathrm{Mag}_{sersic_{\mathrm{true}}}\).  Models selected by Bayes
    factor as containing point sources are indicated in red
    ($\ln\,BF_{12}<-1$). Blue points show galaxies where the Bayes
    factor favors a pure Sersic model ($\ln\,
    BF_{12}>1$). Green points indicate intermediate cases ($\ln\,
    BF_{12}>-1$ and $\ln\,BF_{12}<1$).  The solid lines
    correspond to the posterior median and the shaded area shows the
    1-$\sigma$ quantiles. \label{biasPs}}
\end{figure}

For every image, we need to compute separately the posterior
distribution under the assumptions of both models, $M_{1}$ and
$M_{2}$.  The value of \(BF_{12}\) required to \emph{believe} either
assumption is a matter of cost and risk.  Astronomers, because they
have a finite amount of expensive data, are willing to take on risk.
According to Jeffrey's interpretation, when $\ln\,BF_{12}>1$
($<-1$) the evidence is strongly in favor of $M_{1}$
($M_{2}$). Galaxies in the range $-1<\ln\,BF_{12}<1$ are
considered ambiguous. There are many version of these criteria; see
\citet{RobertRaftery} for another commonly used version.  When we
consider all galaxies without a PS ($M_{1}$) from the ensemble, the BF
correctly prefers model $M_1$ in 67 of 72 cases; 5 cases were
classified as ambiguous and 1 case was incorrectly attributed to model
$M_{2}$. Therefore, the reliability obtained is $93.0\%$ and the
median and dispersion of \(\ln\,BF\) for the true \(M_1\) sample
are $2.27\pm1.02$.  For galaxies with a nuclear point source, the BF
classifies 34 cases of 360 as $M_2$.  If we define the null hypothesis
as \emph{the galaxy does not present a nuclear point source}, false
negatives are below $8.3\%$.  Our ability to recover the point-source
magnitude in an extended source depends on the $\delta
\mathrm{Mag}_{\mathrm{true}}$, the ratio $r_{e}/$FWHM and the S\'ersic
index $n$ (see red points in Figure \ref{biasPs}).  Restricting our
sample to images with $r_{e}\ge 7.92$\,arcsec, for $n\le 6$ and
$\delta \mathrm{Mag}\le 5$ and for $n>6$ and $\delta \mathrm{Mag}\le
3$, the false negative rate is $(2/24)=8.33\%$ and false positive rate
is $(5/36)=13.8\%$. For this experiment, we assume a fixed PSF FWHM of
1.3\,arcsec, typical of the SDSS.

Figure \ref{biasPs} illustrates the bias in the inferred $\delta
\mathrm{Mag}$ and the BF as a function of $\delta
\mathrm{Mag}_{\mathrm{true}}$, $n$, and $r_{e}$. This figure considers
typical SDSS values for the S/N = 450 and $q = 0.7$. Inspecting the
figure rows, one can see that the BF identifies the PS only when
$r_{e}>=7.92$\,arcsec, i.e. an effective radius 6 times larger than
the FWHM of the PSF. Looking along the columns, as $n$ becomes larger
the BF sensitivity decreases.  The PS contribution is negligible when
$\delta \mathrm{Mag}>=5$ and GALPHAT returns a finite large value for
$\mathrm{Mag}_{PS}$ even when there is no PS.  Therefore, the bias in
\(\mathrm{Mag}_{PS}\) decreases as \(\delta \mathrm{Mag}\) increases,
i.e. the estimated PS magnitude becomes fainter (cf. the first row in
Fig. \ref{biasPs}).  Finally, the BF can identify the PS only when
they are bright enough ($\delta \mathrm{Mag} < 5$), and $r_{e}$ is
greater than 7.92\,arcsec. For $n=8$ and $r_{e}=15.84$\,arcsec, we can
see that the BF detects the PS only for $\delta \mathrm{Mag} = 3$ or
brighter.

\begin{table}[!ht]
  \begin{center}
    \caption{The confusion matrix for point-source
      selection\label{conf_matrix}}
    \vskip 0.2 truecm
    \begin{tabular}{lccccc}    
      \hline\hline
      & S\'ersic  & Indeterminate & S\'ersic+Point Source& Total\\
      & $\ln\,BF>1$& $1<\ln\,BF< 1$&  $\ln\,BF < -1$&
      \\\tableline
      S\'ersic  & 21
      &  1& 2&24 \\ 
      S\'ersic +Point Source & 5  & 0 & 31&36\\ 
      \tableline
    \end{tabular}
  \end{center}
\end{table}

To assess the detection and classification errors caused by an
offset of the galaxy center
from the image center (note that the PS is always at the galaxy
center), we created another two samples of synthetic
galaxies with point sources. The first sample has the galaxy center at the
central pixel of the image, while the second has small shifts from the central pixel. Our experiments have shown that when the galaxy center is slightly shifted (by fraction of a pixel), 43 out of 45 galaxies with $\ln\,BF\leq-1$ are still well classified by the BF.

\section{Application to SDSS ETG images}
\label{resultSDSS}

\begin{figure}[!ht]
  \epsscale{0.8}
  \plotone{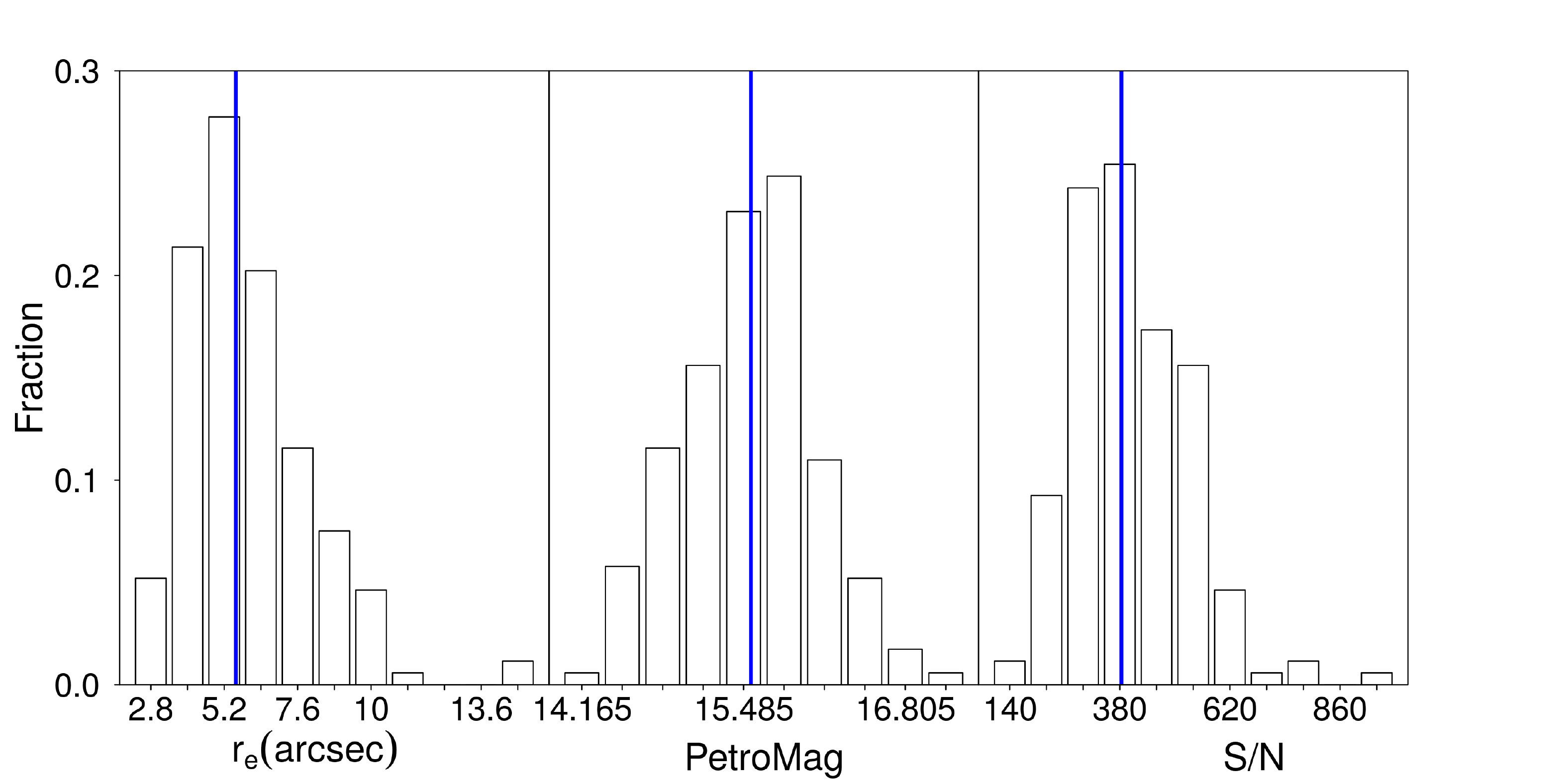}
  \caption{Parameter distribution in our SDSS sample: effective radius
    $r_{e}$ (left), Petrosian magnitude (center), and S/N (right).
    The median values are indicated by the blue vertical
    lines.\label{propertiesSample}}
\end{figure}

We selected early-type galaxies from SDSS DR7 with $\log M_{stellar} >
11.50$, eClass < -0.2 in the redshift range of 0.05 to 0.1 classified
as \emph{Elliptical} by the Galaxy Zoo 1 project \citep{zoo2011}. These
are very luminous, high S/N, systems.  The threshold in eClass
(morphological classification based on the spectrum) usually is set to
zero to discriminate early from late type galaxies, thus setting it to
-0.2 minimizes even further any possible contamination from late-type galaxies
that may be left from the Galaxy Zoo classification.  The minimum redshift
reduces aperture effects in the measurements required for the AGN
diagnostics (presented in Section \ref{sec:AGN}) and the upper limit
ensures spectroscopic completeness of the SDSS-DR7. The resulting
sample of 200 bright ETGs have all been processed through the SDSS
imaging pipeline which fits two models to the two-dimensional image of
each object in each band: (i) a pure de Vaucouleurs profile and (ii) a
pure exponential profile. Figure \ref{propertiesSample} shows the
effective radius $r_{e}$, the Petrosian magnitudes, and S/N
distribution for the sample.  We use the catalog values to define the
\emph{postage stamp} images and to specify our prior distributions (as
described in Sections \ref{prior} and \ref{settingGALPHAT} and Table
\ref{priorsTable}).  The stamp size was chosen such that the galaxy center is 7.5 de Vaucouleurs effective radii from the stamp edges.  We further classify the
quality of each image by a quality flag (QF), in Table \ref{tab:QF}.

GALPHAT produced converged posterior distributions for 166 galaxies
out the 200. A detailed inspection of the log files, residuals and
posteriors in the 34 unconverged cases indicates that most of them are
close to the frame edges (QF=3) and have a secondary object covering the
central region (QF=1).  All galaxies with QF=0 resulted in converged
posterior distributions.

\begin{table}[!htb]
  \caption{Postage-stamp quality flags}
  \label{tab:QF}
  \centering
  \begin{tabular}{lll}
    QF & Description & Count \\ \hline
    0  & high quality & 33 \\
    1  & close to frame border & 31 \\
    2  & some secondary sources in the outer region & 132 \\
    3  & some secondary sources in the central region & 4 \\
  \end{tabular}
\end{table}

We now select a 102 ETG subsample with $r_{e}>7.92$ arcsec (our
\emph{safe} detection limit according to Sec. \ref{recoverPS}) and
combine 10000 randomly-selected converged states from the posterior and
illustrate the inherent covariance between models (S\'ersic and
S\'ersic plus Point Source) parameters.  Using the combined posterior
distributions, rather than best-fit ML solutions, we naturally include
the intrinsic correlations in the parametric model induced by the
data.  That is, the posterior distribution will be broadened in the
parameter coordinates that suffer intrinsic correlations.  Using the
posterior sample, rather than a biased single point estimate, allows
interesting, subtle details about the entire population to emerge.

\begin{figure}[!ht]
  \mbox{
    \includegraphics[width=0.32\textwidth]{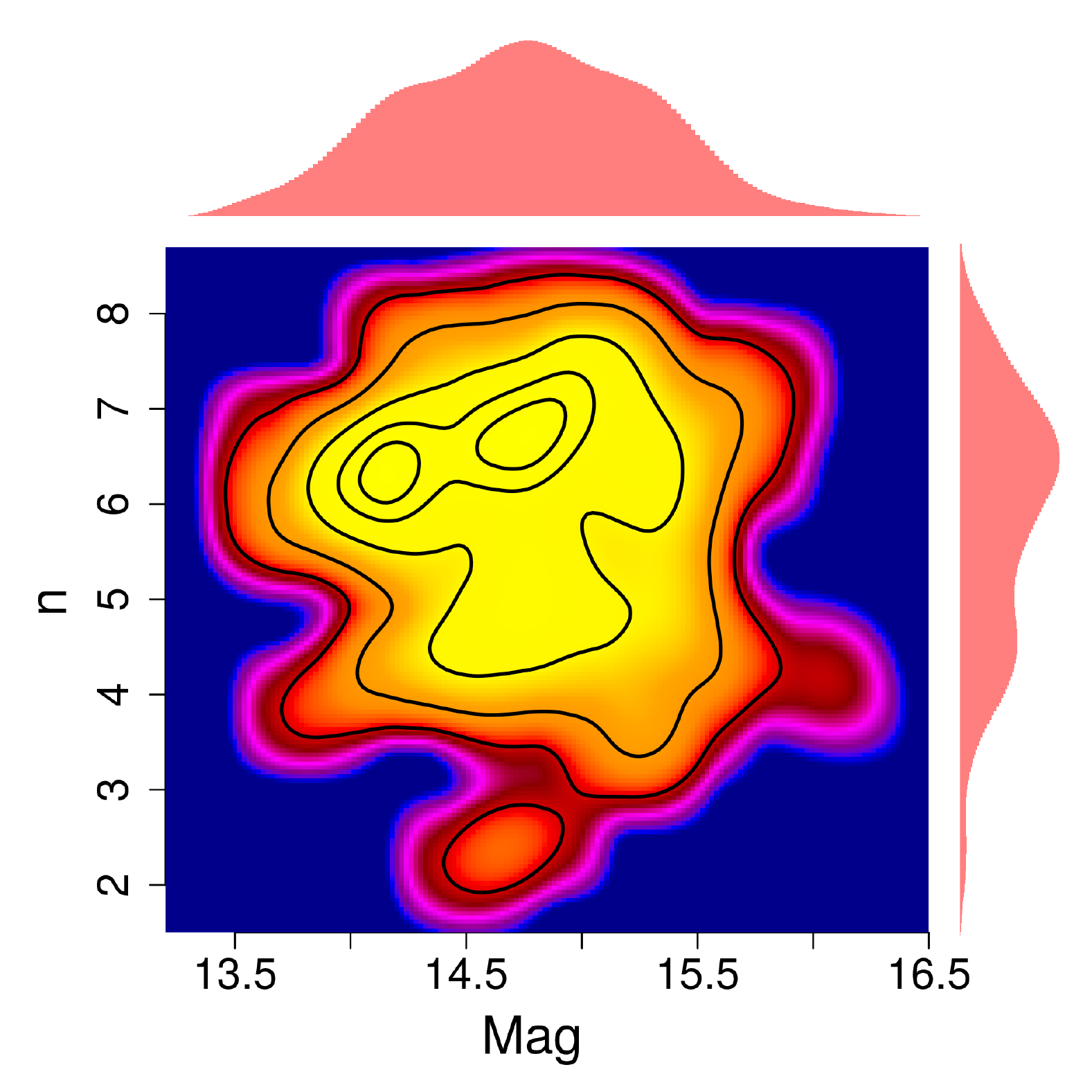}
    \includegraphics[width=0.32\textwidth]{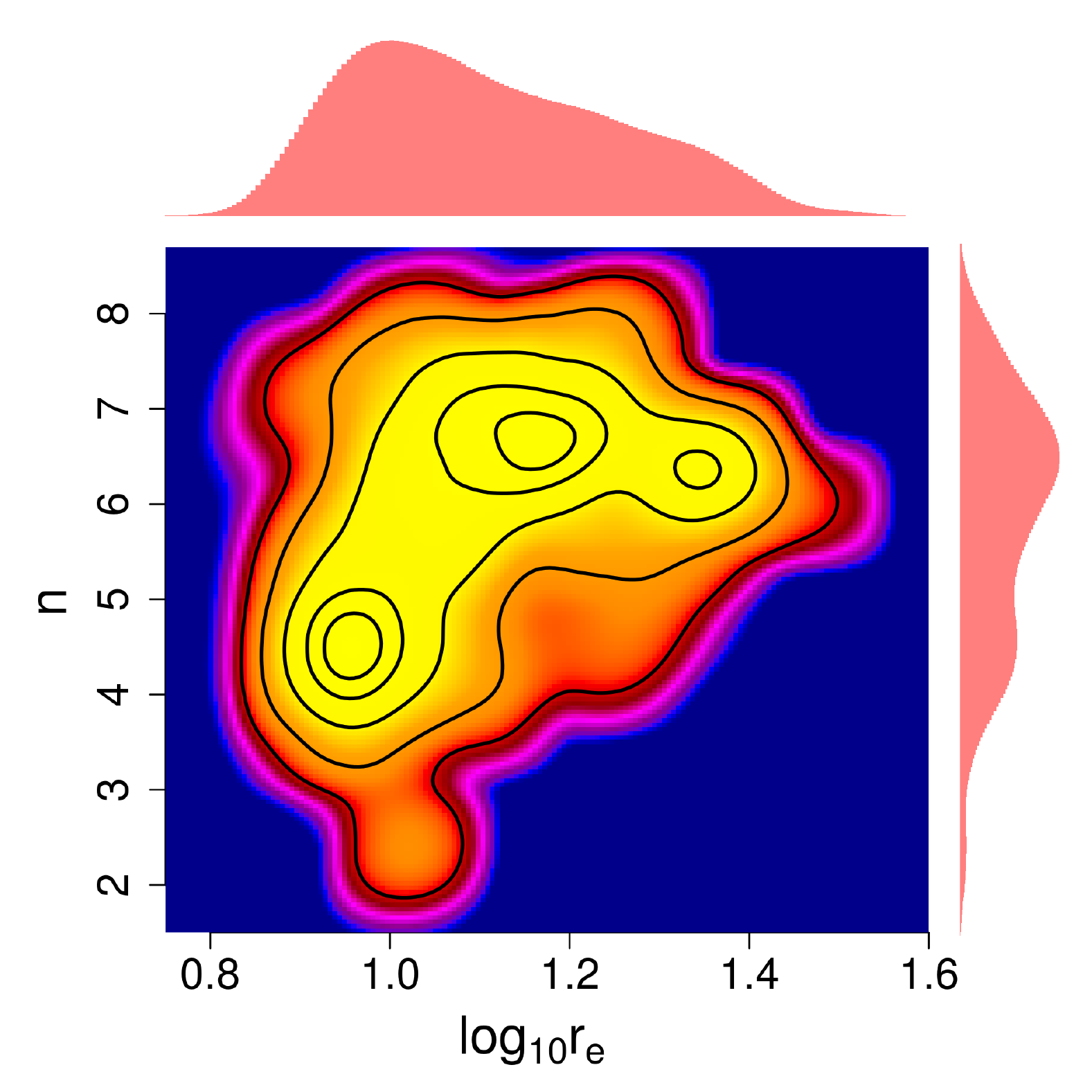}
    \includegraphics[width=0.32\textwidth]{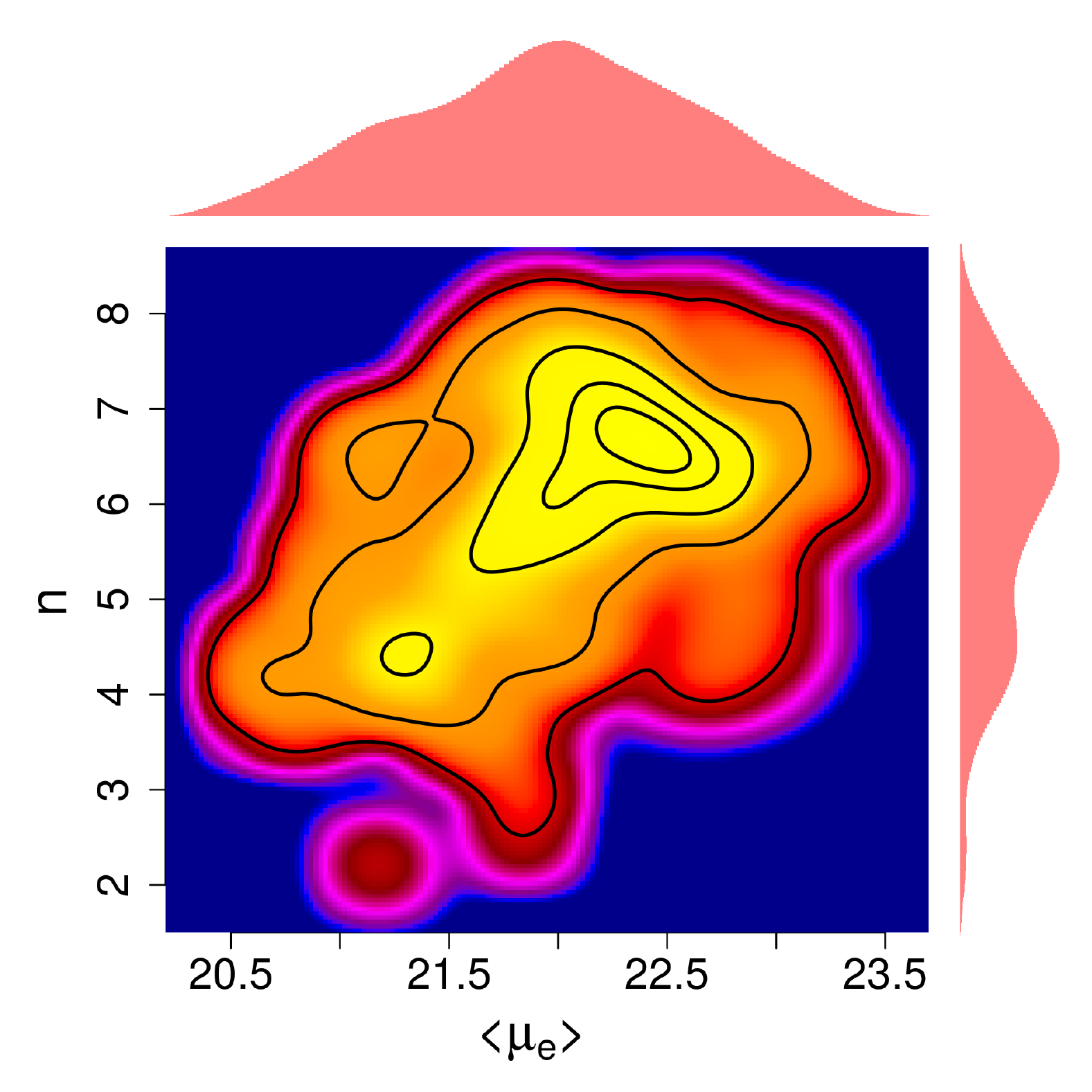}
  }
  \caption{Two dimensional distributions of the combined posterior
    for galaxies with \(r_{e}>7.92\)\,arcsec and  \(\ln\, BF\geq1\) and \(\ln\,
    BF\leq-1\) in our SDSS
    EGT sample. We combine samples from each posterior weighting by the 
    probability odds. 
    \label{covariances}}
\end{figure}
 We combine the information provided by both models (S\'ersic and
S\'ersic plus Point Source), weighting samples from each posterior by the
probability odds, e.g. $w_{i:1,2}=\frac{P(D|M_{i})}{P(D|M_{1})+P(D|M_{2})}$.
Figure \ref{covariances} shows two-dimensional distributions for 
various S\'ersic-model parameters. The middle panel illustrates the
correlation between the S\'ersic index and the effective radius and
reveals a distinct bimodal distribution. The first mode, $M_{A}$,
peaks at n$\sim$4.5 and log $r_{e} \sim$0.95, and the second, $M_{B}$,
peaks at n$\sim$6.8 and log $r_{e} \sim$1.15. These two peaks seem to also 
be present in the other two projections, versus magnitude and
mean surface brightness.
The mean surface brightness at $M_{A}$ is significantly higher than at $M_{B}$.
This lends further credibility that the bimodal nature of the
\(n\) distribution is real. 
\subsection{AGN classification}
\label{sec:AGN}

We now correlate our 102 ETG subsample with a
spectroscopic AGN signature.  Our spectroscopic analysis is motivated
by the WHAN diagram \citep{Cidetal05}. This diagnostic diagram uses
the $\log [NII]6584/H\alpha$ ratio and the equivalent width of the
$H_\alpha$ line. The main advantage of the WHAN diagram over other
diagnostic diagrams (like the \citet{BaldwinEtal1981} diagram) is the
ability to separate LINERs from ``fake-AGN'' -- galaxies with
low-ionization emission spectra that resemble LINERs but are not caused by
nuclear activity. It is, therefore, a useful tool to discriminate
between galaxies whose main ionization mechanism owes to
main-sequence high-mass stars (i.e. star-forming galaxies), galaxies
where the gas is ionized by active galactic nuclei (Seyfert-like or
LINER spectra), LINERs or ``retired'' galaxies (i.e. galaxies whose
emission lines are produced by ionization from hot evolved low-mass stars, HOLMES), and passive galaxies.

Using the CASJOBS
interface\footnote{http://skyserver.sdss.org/casjobs/}, we have
obtained the equivalent widths and line fluxes of $H\alpha$ and
$[NII]6584$ from \citet{Thomasetal2013}, derived from the SDSS-III
single fiber spectra. These data were available for only 94 ETGs from
our $r_e>7.92$ arcsec sample.  Our \emph{safe} limit assumes a PSF
FWHM of 1.3 arcsec. 

We find that all galaxies in this subsample have $\log
[NII]6584/H\alpha>-0.4$. We conclude, therefore, that there are no
star-forming galaxies in this sample. This is not surprising given
that our sample is composed of massive elliptical galaxies that in
general lack active star formation. In the WHAN diagram, the
ionization level of non-star forming galaxies is indicated by the
equivalent width of the H$\alpha$ line ($W_{H\alpha}$). An ionization
spectrum dominated by active galactic nuclei corresponds to $\log
W_{H\alpha}>0.48$, while a passive or \emph{lineless} galaxy is
defined by $\log W_{H\alpha}<-0.52$ and $W_{[NII]}<-0.52$;
intermediate values correspond to retired galaxies. Notice that, as
shown by \citet{Cidetal05}, there is a considerable scatter around
these limiting values.

\begin{figure}[!ht] 
  \centering 
  \includegraphics[width=0.8\textwidth]{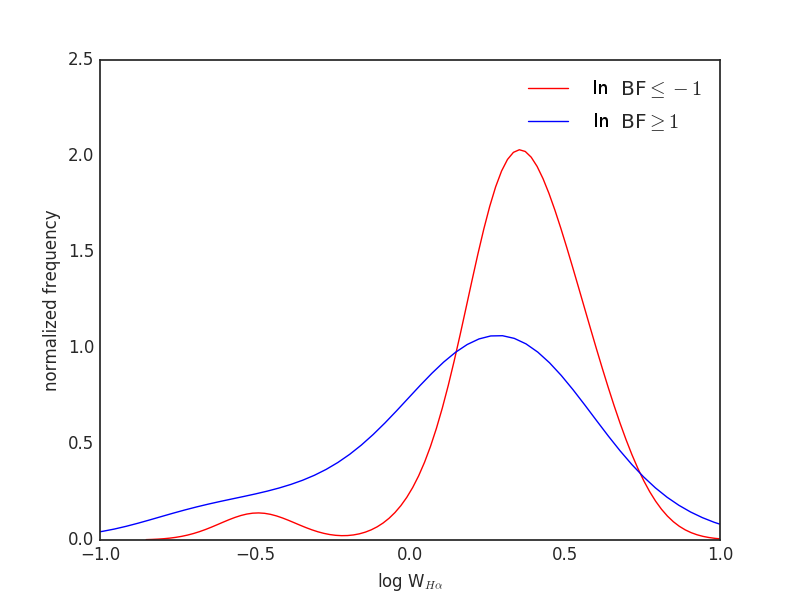}
  \caption{Distributions of the H$\alpha$ equivalent widths (in \AA)
    for a subsample of ETGs with $r_{e}$>7.92 arcsec. The blue and red lines represent
    galaxies in the ranges \(\ln\, BF\geq1\), and \(\ln\,BF\leq-1\) respectively. A
    Gaussian density kernel has been applied to the discrete data.}
  \label{BFdistribution}
\end{figure}

Figure \ref{BFdistribution} presents the distribution of $W_{H\alpha}$
for our galaxy subsample; a Gaussian density kernel has been applied to
the discrete data. We have divided the subsample into three ranges of BF
according to the model that best describes the surface brightness
distribution: objects for which a S\'ersic+PS is favored
(\(\ln\,BF\le-1\)), and objects for which a single S\'ersic
profile is favored (\(\ln\,BF>1\)). Only one galaxy was located in
the intermediate region where both models have similar probabilities,
and was therefore discarded from further analysis. We can see that the
distributions of $W_{H\alpha}$ are markedly distinct for the two BF
ranges. Galaxies with \(\ln\,BF\leq-1\), i.e. galaxies for which a
S\'ersic+PS model is favored, are concentrated around the dividing
line between LINERs and retired galaxies, at $\log
W_{H\alpha}\sim 0.35$. In contrast, galaxies for which GALPHAT does
not indicate the presence of a nuclear point source (\(\ln\,
BF\geq1)\) are distributed across the full range of the
LINER--retired--passive regions. The Anderson-Darling test indicates
that the distributions of $\log W_{H\alpha}$ for galaxies with $\ln\,\leq-1$ and $\ln\,BF\geq1$ are dissimilar at the 98.6\% level.

Low-ionization emission-line regions are present in both AGN,
star-forming and non-star-forming galaxies.  All of these lead to
detectable \(H\alpha\).  The distribution of \(H\alpha\) equivalent
width in passive, non-starforming galaxies is broad and extends to low
values; the equivalent width in galaxies hosting a modest central
emission source is generally larger than approximately 1 \AA\, 
\citep[e.g.][]{Belfiore.etal:2016}.  In our sample, there are
elliptical galaxies with residual hydrogen being ionized by a number
of different sources, be it nuclear or extended in the galaxy.  The
WHAN diagram suggests than no objects in our sample are star-forming.
Therefore, the distribution of \(W_{H\alpha}\) is expected to be broad
and extending to low values of \(W_{H\alpha}\) for galaxies identified
as single S\'ersic profile by GALPHAT.  Conversely, even underluminous
active nuclei will provide enough ionizing photons to result in a lower floor
for WHa.  Therefore, the tight distribution for
\(\ln\,BF\leq-1\) and broad distributions for \(\ln\,BF\geq1\)
are consistent with our photometric identification of central sources.

In summary, Bayes-factor model selection using GALPHAT identifies
central point sources when the \(W_{H\alpha}\) is in the vicinity of
the LINER/retired dividing line. Conversely, profiles without central
point sources (large values of BF) have a broadly extended
\(W_{H\alpha}\) distribution, consistent with passive LINERs.  Also,
GALPHAT does not detect point sources for a significant number of
emission-line galaxies where the emission is expected to be extended.
This is further evidence that GALPHAT is properly discriminating between
galaxies with and without unresolved nuclear emission.

\section{Summary}\label{summary}

We explore the Bayesian inference of the structural parameters for a
population of early-type galaxies (ETGs) using the BIE
\citep{Bie2013t} and GALPHAT.  We quantify the accuracy and
reliability for extremes in the ETG population and the classification
of AGN photometric morphology using Bayes factors.  This work
introduces a Python-based pipeline, PyPiGALPHAT\footnote{ PyPiGALPHAT
  source code is available on request at
  \url{git@bitbucket.org:diegostalder/pypigalphat.git.}}  for
automating and documenting the production posterior distributions on
high-performance computing clusters.  Our key findings are as follows:
\begin{itemize}
\item Using simulated images tuned to SDSS images based on S\'ersic
  profiles, we benchmarked GALPHAT for high, \(4<n<10\), S\'ersic
  index values for the expected ranges of PSF widths and
  signal-to-noise ratios.  The bias in the inference of structural
  parameters is larger for a concentrated distribution (\(n\ge8\)).  Our
  tests extended the parameter space range of the initial benchmarks
  done by YMK10.
\item A comparison between the GALPHAT posterior distribution and
  GALFIT ML estimates reveals negligible biases for S\'ersic \(n\le2\) in both
  methods.  The
  GALFIT bias is significantly larger than GALPHAT bias for higher $n$
  values. For extreme values, e.g. $n = 10$, the GALPHAT bias for $n$
  is at least three times lower than the GALFIT bias. The bias for $n$
  is positive, raising the concern that GALFIT can lead to
  significantly overestimated S\'ersic indexes.
\item The BIE efficiently estimates the marginal likelihood by
  resampling \citep[e.g.][]{Weinberg2012}, enabling Bayes factor model
  comparison.  We tested the power in Bayes factor selection between
  pure S\'ersic profiles and S\'ersic profiles with central point
  sources.  The BF can reliably identify central point sources of
  galaxies with effective radii larger than $7.92$\,arcsec PSF FWHM and
  pixel scales typical of SDSS.  For low (high) S\'esic indexes $n<=6$
  ($n> 6$), we can identify point sources with magnitudes 5 (3) mag
  fainter than the extended galaxy. We find that false positive and
  negative classification errors are below \(14\%\).  For a PSF and
  pixel scale typical of HST, central point sources can be identified
  in galaxies with a typical effective radii of \(3.96\)\,arcsec and
  \(\delta\mathrm{Mag}\lesssim 5\).
\item The combined posterior density for a well-defined galaxy sample
  allows us to characterize trends and features that describe that
  population.  For example, we show that our approach automatically
  reveals a bimodal distribution.  The first mode, $M_{A}$, peaks at a S\'ersic of index of \(n\approx 4.5\) and the second mode,
  $M_{B}$, peaks at an index of \(n\approx6.8\).  In addition, mode
  $M_{B}$ is characterized by a larger effective radius.
  \item Our Bayes-factor classified point-source detections correlate
    with AGN spectral signatures.  The members of our ETG sample are
    distributed between the LINER and retired regions of the WHAN
    diagram. Our pure S\'ersic sample has a much broader distribution
    in \(H\alpha\) equivalent width, going all the way from LINER to
    retired to passive galaxies. On the other hand, galaxies with
    point sources are concentrated in the region that separates the
    LINER and retired regions, as expected for a true central emission
    source. We show also that the observational condition
    characterized $r_{e}$/PSF FWHM can significantly affect the
    reliability of the analysis.
\end{itemize} 

We have demonstrated that GALPHAT offers several important advantages
over other commonly used 2-d galaxy photometry fitting codes.  First,
GALPHAT produces the full posterior distribution, not just the maximum
likelihood solution, which reduces biases in correlation estimates as
demonstrated in \citet{GALPHAT2013}.  This allows the trivial
computation of credible intervals for any parameter.  Secondly,
GALPHAT's rigorous internal error control allows GALPHAT to robustly
determine parameter fits without the need to mask out the central
regions, as is often required when using GALFIT.  The increased
accuracy, reduced bias, and our Bayesian approach allows us to
simultaneously fit for the background when we fit the galaxy.  This
facilitates low-bias parameter estimations since the background
uncertainty can be strongly covariant with other parameters in many
cases \citep{2007GEMS,2009GUO,2016Robotham}. GALPHAT also gives robust
and unbiased parameter estimations down to signal to noise ratios of a
few, enabling parameter estimation and classification for faint
galaxies in surveys like CANDELS \citep{CANDELS1:2011,
  CANDELS2:2011}. Finally, by calculating marginal likelihoods and
Bayes ratios, GALPHAT naturally presents the odds that one model is
favored over another. For example, in both GALPHAT and GALFIT one can
fit the galaxy image either using one Sersic component, a Sersic
component plus an exponential disk component, or either model with a
central point source, but GALPHAT computes the relative probability of
the models in light of the data, e.g. using the Bayes ratio.

Our tests have demonstrated that we can achieve a steady-state
posterior distribution in a wide range of typical astronomical regimes
and that the simulated posterior will include all multiple modes
consistent with the prior distribution.  Using the posterior
distribution, we show that the surface-brightness model will often
have correlated parameters and, therefore, any hypothesis testing that
uses the ensemble of posterior information will be affected by these
correlations.  Our results suggest that Bayesian photometric analysis
has the ability to discriminate between competing models.  In
particular, we have demonstrated with low-resolution SDSS data that we
can reliably detect AGN photometric signatures as long as the PSF is
smaller than to the characteristic scale of the extended light
profile.  This method will excel at characterizing high-resolution images
from upcoming high-resolution ground based and space-based facilities
(e.g. TMW, VLT, and JWST).

\acknowledgments \textbf{Acknowledgments:} DHS acknowledges the
financial support from CNPQ scholarship $140913/2013-0$. The authors
acknowledge financial support from FAPESP through a grant
$2014/11156-4$. S. B. R. acknowledges support from FAPERGS. Much of the work presented here was made possible by
the free and open R software environment (R Development Core Team
2016). The authors thank to MCTIC/FINEP/CT-INFRA project (grants
0112052700) and the Embrace Program.

\appendix

\section{Performance: Data Management and Run time}
\label{performancesec}

Large input data sets, intermediate data products, configuration
files, and posteriors distributions require organization for
successful handling. At each step, the pipeline maintains a list of
files that must be saved or deleted.  The most important files and
their sizes are listed below:
\begin{enumerate}
\item Pre-processing: the input FITS image data (the full frame,
  postage stamps, and masks; 5 MB), configuration files.
\item Processing output: the posterior samples (ascii; 50 MB), MAP and
  ML image residuals (50 MB), simulation persistence data (1GB), and
  log files.
  \begin{enumerate}
  \item[2.2] Pre-postprocessing: compressed posterior samples
    (FITS.gz; 5 MB), image residuals (7 MB), and log files.
  \end{enumerate}
\item Postprocessing: the marginal (140 KB) and posterior distribution
  plots (155 KB), cumulative covariance data (1.4 MB), residual png
  images (420 KB), and output catalog having the pooled posterior and
  inferred figures of merit like the MAP, median, and variances
  etc.
\end{enumerate}
Table \ref{data_demans} shows a summary of disk space that we need for
each stage of the pipeline for our SDSS sample of 200 ETGs.
\begin{table}[ht]
  \begin{center}
    \caption{Data generated on each stage of the
      pipeline\label{data_demans}}
    \vskip 0.2 truecm
    \begin{tabular}{@{} *2l @{}}    
      \hline\hline
      \textit{Step} & Disk Space \\
      &  (200 Galaxies)
      \\ \tableline
      Data Frames &  1 TB   \\ 
      Preprocessing &  350 MB   \\ 
      Processing  &  2.5 TB\\ 
      Pre-Post-Processing  &
      1.5 GB \\ 
      Post-Processing & 750 MB  
      \\
      \tableline
    \end{tabular}
  \end{center}
\end{table}

\begin{figure}[ht]
  \centering
  \includegraphics[width=0.8\textwidth]{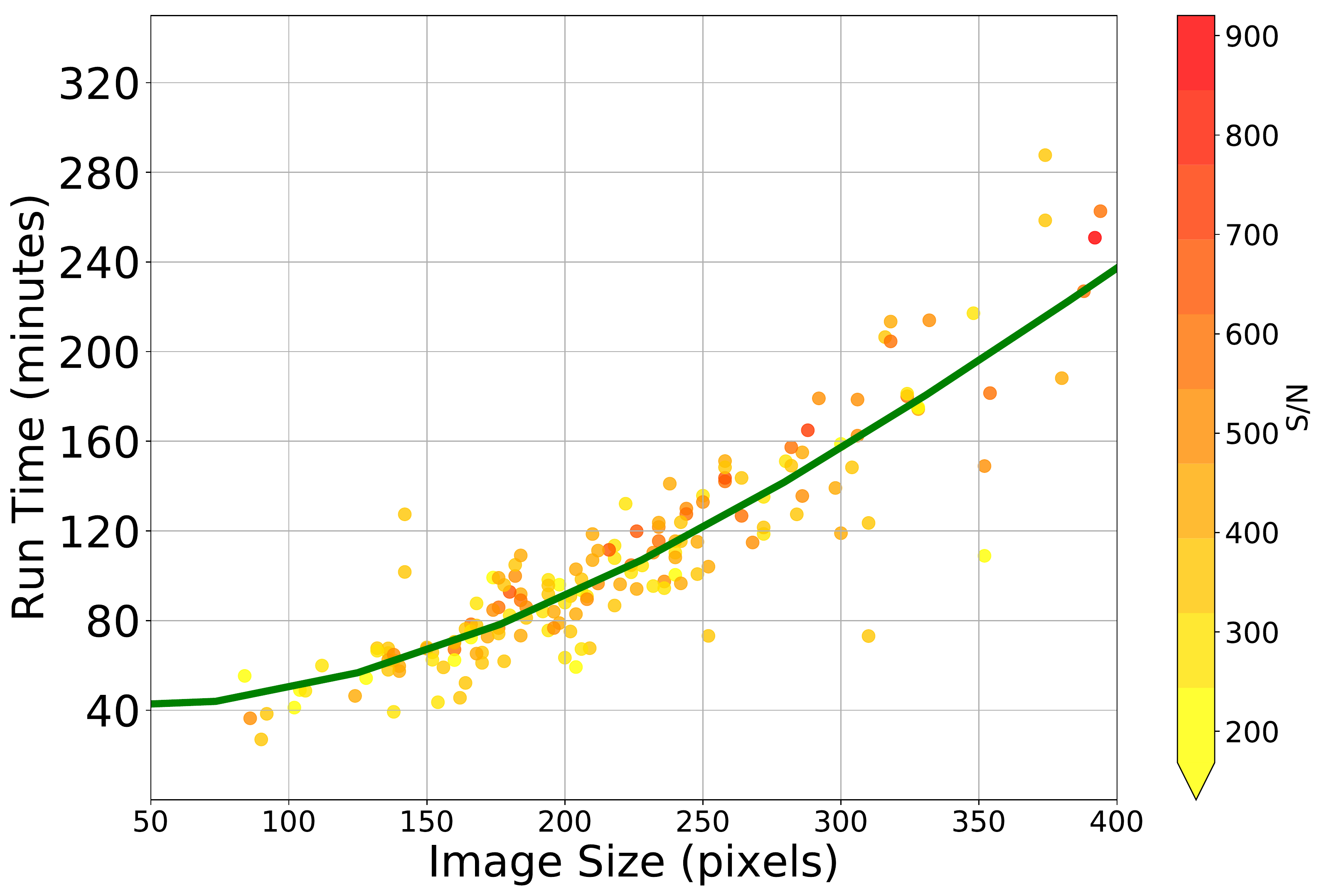} 
  \caption{Total GALPHAT run time for each galaxy of our SDSS
    sample. The point colors scale indicate the S/N measured for each
    galaxy. The green solid lines show a linear least-squares
    fit.\label{runTimeTest}}
\end{figure}

Figure \ref{runTimeTest} describes the total run time of one MCMC
simulation and the likelihood marginalization that is used to compute
the Bayes factors as function of the $S/N$ and postage stamp size (in
pixels). The total run time was computed using 10 nodes of our CPU
cluster. A least square fitted curve is: $-1.8 \times 10^{-6}x^{3} +
2.326 \times 10^{-3}x^{2} - 0.1591 x + 43.79 $

\section{PyPiGALPHAT}
\label{pypigalphatsec}

The PyPiGALPHAT pipeline is implemented in Python, csh shell, and R. We
developed a set of routines and scripts to retrieve the galaxy images
from servers, produce the \emph{postage-stamp} images, identify
contaminating sources, generate masks, run GALPHAT, and analyze the
data products. The pipeline has three modules: (i) pre-processing (ii)
processing and (iii) post-processing. Table \ref{PypigalphatModules}
summarizes the function of each module.  The details of the modules
will be described in the following sections.

\begin{table}[ht]
  \begin{center}
    \caption{PyPiGALPHAT modules\label{PypigalphatModules}}
    \vskip 0.2 truecm
    \begin{tabular}{@{} ll @{}}    
      \hline\hline
			
      Modules & Stages \\
      \tableline
      Preprocessing  &  Retrieve data   \\ 
      &  Produce image stamps and masks   \\ 
      &  Produce GALPHAT input file and script \\ 
      Processing  &  Run GALPHAT on HPC hardware\\ 
      Postprocessing  & Quick diagnosis Images\\ 
      & Generate output catalog  
      \\
      \tableline
    \end{tabular}
  \end{center}
\end{table}

\subsection{Pre-processing: fetching postage-stamp images, producing
  image masks and generating GALPHAT input files}
\label{subSecPreprocessing}

The main preparatory steps done by PyPiGALPHAT before estimating the
structural parameters using GALPHAT are as follows: (1) build a list
with the sky coordinates of the target galaxies; (2) retrieve galaxy
metadata from the survey database; (3) download images and PSFs from
the survey image server; (4) produce image stamps containing the
target galaxies; (5) detect objects in the image stamps and generate
preliminary photometric parameters by using SExtractor; (6) generate
image masks to exclude non-target features (foreground and background
galaxies, diffraction spikes, etc.); (7) produce GALPHAT input files
and scripts to obtain structural parameters for a given model
(S\'ersic, S\'ersic plus point source and S\'ersic bulge plus
exponential disk); (8) classify the images to produce a quality flag
(QF, see Section \ref{resultSDSS}). Many of these details are survey
independent and, therefore, minimal changes will be necessary to apply
PyPIGALPHAT to another survey. Additionally, PyPiGALPHAT is designed
for production with simulated images by simply skipping Steps (2) and
(3).

\subsubsection{Retrieving SDSS Data}

We begin by tabulating the exact location of each galaxy, the desired
the photometric band (\texttt{RA, DEC, band}). From this list, the
pipeline builds SQL queries to retrieve information from the survey
data base. Specifically, for the SDSS, we build unique combinations
of \texttt{ObjIDs, run, rerun, camcol, field}.  Then, we download the
required data files: (i) images having 2048 $\times$ 1490 pixels
obtained by the photometric data stream from each CCD; (ii)
\texttt{psFields} which is used to extract the PSF; and (iii)
\texttt{tsFields} which contains the statistics of the photometric
pipeline of
SDSS \footnote{\url{http://www.astro.princeton.edu/PBOOK/datasys/datasys.htm\#astropip}}.

The generated SQL query also retrieves a list of photometric
parameters (\texttt{petroMag, petroMagErr, rowc, colc, deVRad, deVAB,
  deVPhi}) from the SDSS imaging pipeline
\citep{2001Lupton,2002SDSSpipeline}. This pipeline has been used to
analyze the raw telescope images, produce calibrated FITS files, and
build catalogs. The main photometric catalog (PHOTO) contains a
large number of measured parameters and uncertainties, like the
structural parameters assuming a de Vaucouleurs profile.

The PSF spatial variations in SDSS are modeled by the Karhunen-Lo\`eve
transform \citep{2001Lupton}.  The data file \texttt{psFields} has all
the information needed to reconstruct the PSF at a desired point in
the frame (\texttt{rowc},\texttt{colc}).  A stand-alone code is
available to recover the
PSF \footnote{\texttt{http://classic.sdss.org/dr7/products/images/read\_psf.html}}
as an unsigned short FITS file whose background level is set to a
standard soft bias of 1000. PyPiGALPHAT removes this soft bias and
estimates the PSF FWHM using the function \texttt{curve\_fit} from
\texttt{scipy}. Finally, the image stamp and PSF FITS headers are updated
with the astrometry and relevant frame keywords.

\subsubsection{Generating postage-stamp images and masks}

The large frames downloaded from the SDSS servers contain multiple
objects.  During pre-processing, the pipeline script selects a section
of the original frame around the target galaxy, producing a
\emph{postage stamp}. Each stamp must contain enough pixels to allow
for a good estimate of sky background fluctuations beyond the
influence of the astronomical source. On the other hand, large stamps
increase computational resource requirements. \cite{2007GEMS} have
shown that sky estimation is of critical importance to correctly
derive the light profiles of galaxies. After several tests and visual
inspection of the output images, a linear size of \texttt{15 devRad}
per side emerged as a good compromise, where \texttt{devRad} is the
effective radius produced by the SDSS photometric pipeline assuming a
pure de Vaucouleurs law.

The steps required for the image stamp production are as follows: (i)
cut out a preliminary stamp with side length of \texttt{17 devRad}; (ii)
identify large objects in the stamp, considering a high detection
threshold (more details are given below); (iii) determine the S/N
ratio using the isophotal flux (\texttt{FLUX\_ISO}) and its RMS error
(\texttt{FLUXERR\_ISO}); (iv) estimate the sky background (SKY); (v)
trim the preliminary image to the final stamp size of \texttt{15
  devRad}; (vi) identify small objects, considering a lower detection
threshold(more details are given below); (vii) extract the information
necessary to compute the calibrated flux (\texttt{zeropoint, airmass},
\texttt{extinction coefficient}, \texttt{gain, readout noise}) from
\texttt{tsFields} data
files\footnote{http://classic.sdss.org/dr7/algorithms/fluxcal.html};
(viii) generate the mask images to avoid non-target objects; (ix)
classify the quality of the stamps by generating stamp quality flags
(SQ) to identify unusual cases.

\begin{figure}[ht]
  \epsscale{.95}
  \plottwo{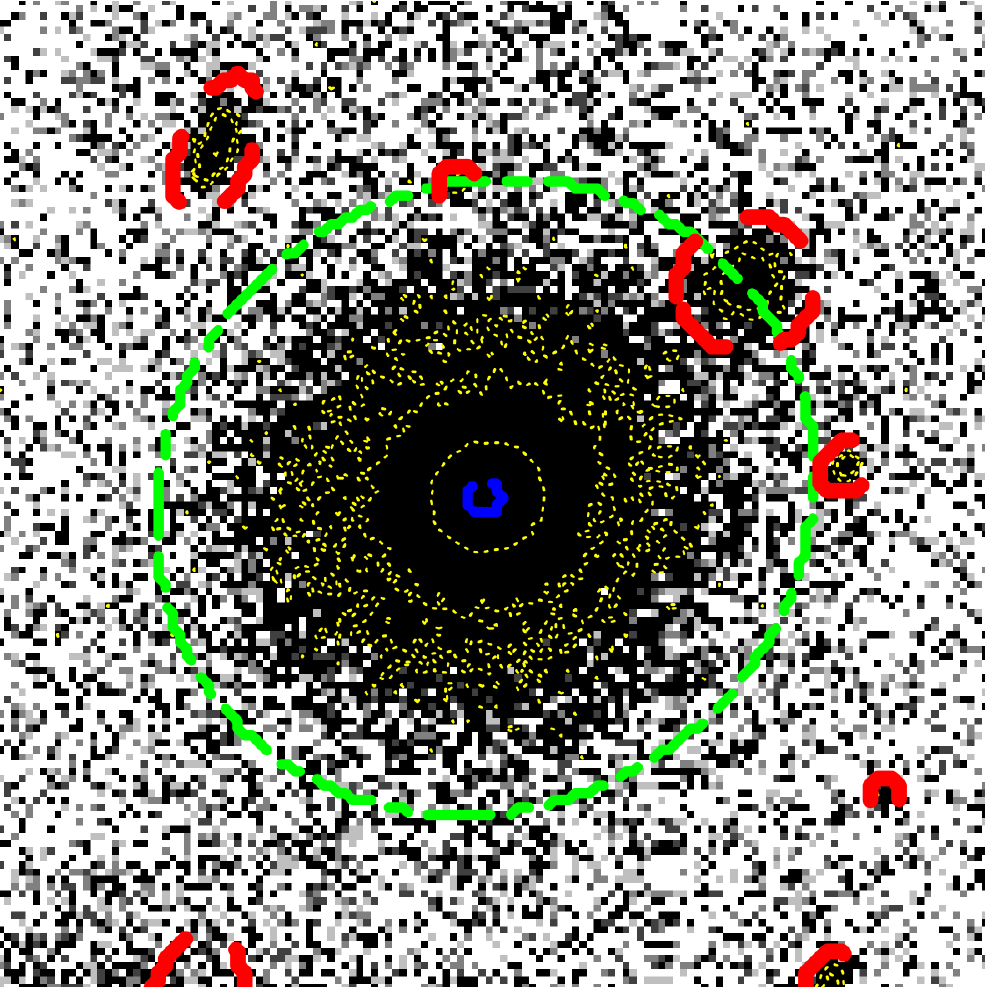}{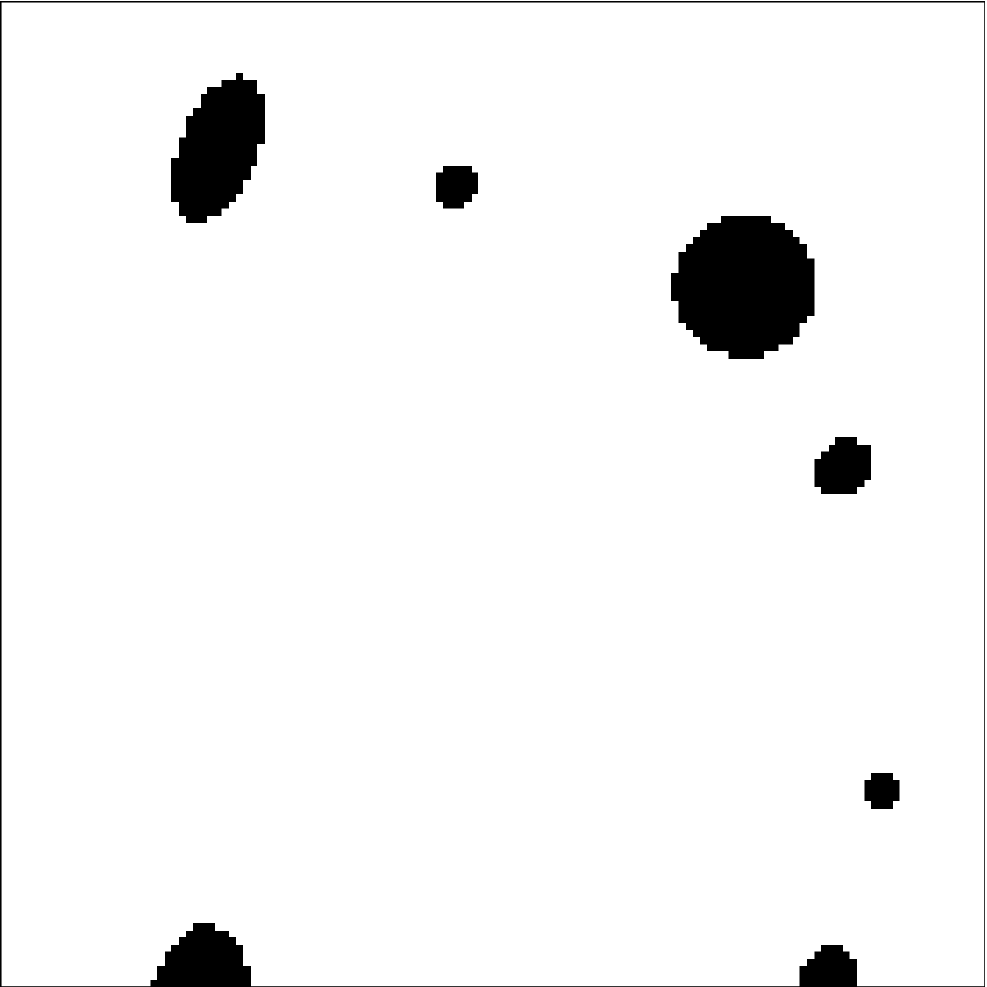}
  \caption{Left (Right) panel shows a typical SDSS diagnostic Stamp
    (mask) obtained by the PyPiGALPHAT preprocessing step. Dotted light
    green contours indicate the target source. Dotted red lines show
    mask objects.\label{stamps2}}
\end{figure}

Each stamp can have photons coming from different objects plus
background fluctuations. An accurate estimate of the background level is
needed to detect the faintest of these objects. PyPiGALPHAT uses
SExtractor \citep{BertinEtal1996} to identify these objects in the
stamp. The detection threshold is controlled directly by
\texttt{DETECT\_THRESH}, \texttt{DETECT\_MIN\_AREA},
\texttt{DETECT\_MAXAREA}. For this work, we use the values 1.3, 3.0
and NONE, respectively. Thus, a new source is tagged as independent if
has it has a flux larger than 1.3 times the standard deviation above the local
background and its area is larger than 3.0 pixels. The local
background estimate depends on the mesh size
(\texttt{BACK\_SIZE}). PyPiGALPHAT controls this detection threshold
indirectly by modifying the mesh size. To detect the larger sources
and accurately estimate the background, we set
\texttt{BACK\_SIZE=100}. On the other hand, to identify small sources
we consider a finer mesh by setting \texttt{BACK\_SIZE=10}.
PyPiGALPHAT creates mask images from the resulting source list.  The
masked area is a combination of ellipses centered at the position of
each secondary object with centers, axis ratios, and position angles
from SExtractor.  The axes for each ellipse are scaled by
\(3\times\texttt{PETRO\_RADIUS}\times\texttt{A\_IMAGE}
(\texttt{B\_IMAGE})\). Figure \ref{stamps2} shows a typical stamp and
mask produced by the pipeline from an SDSS frame.

\begin{figure}[ht]
  \epsscale{0.45}
  \plotone{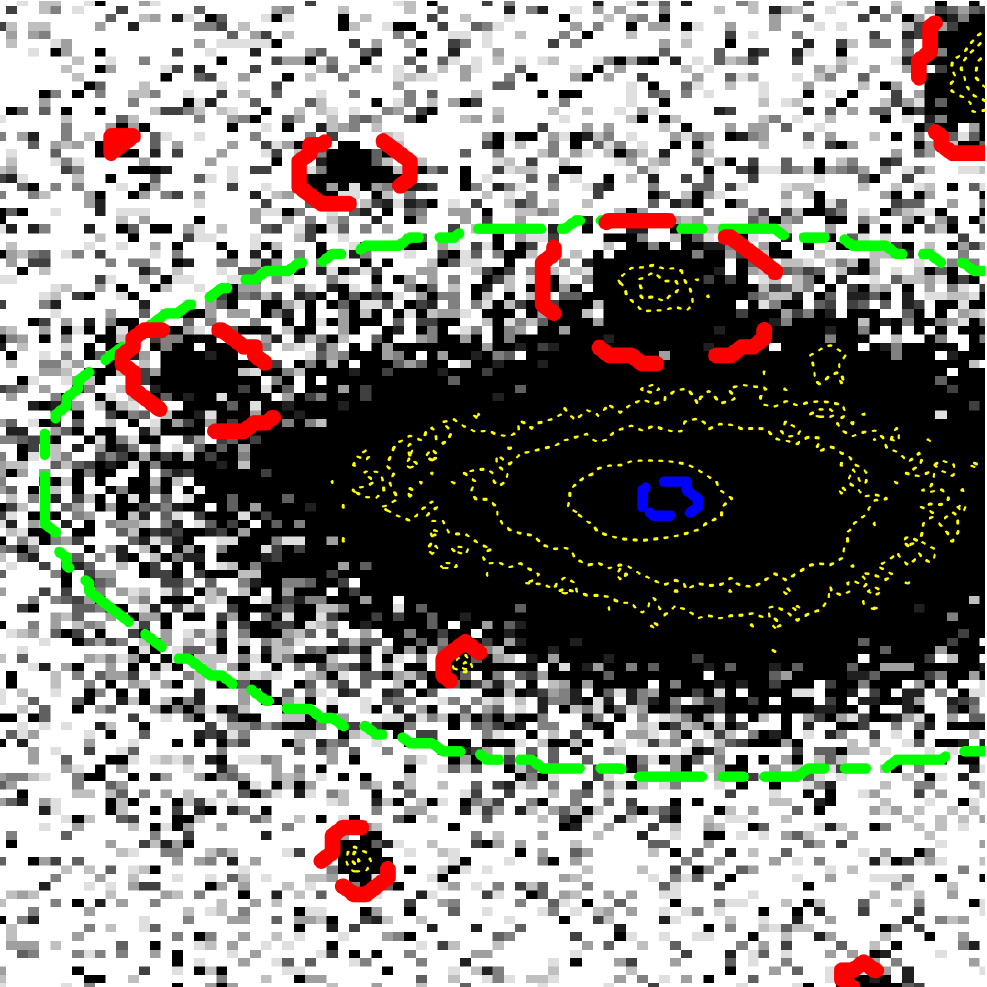}
  \plotone{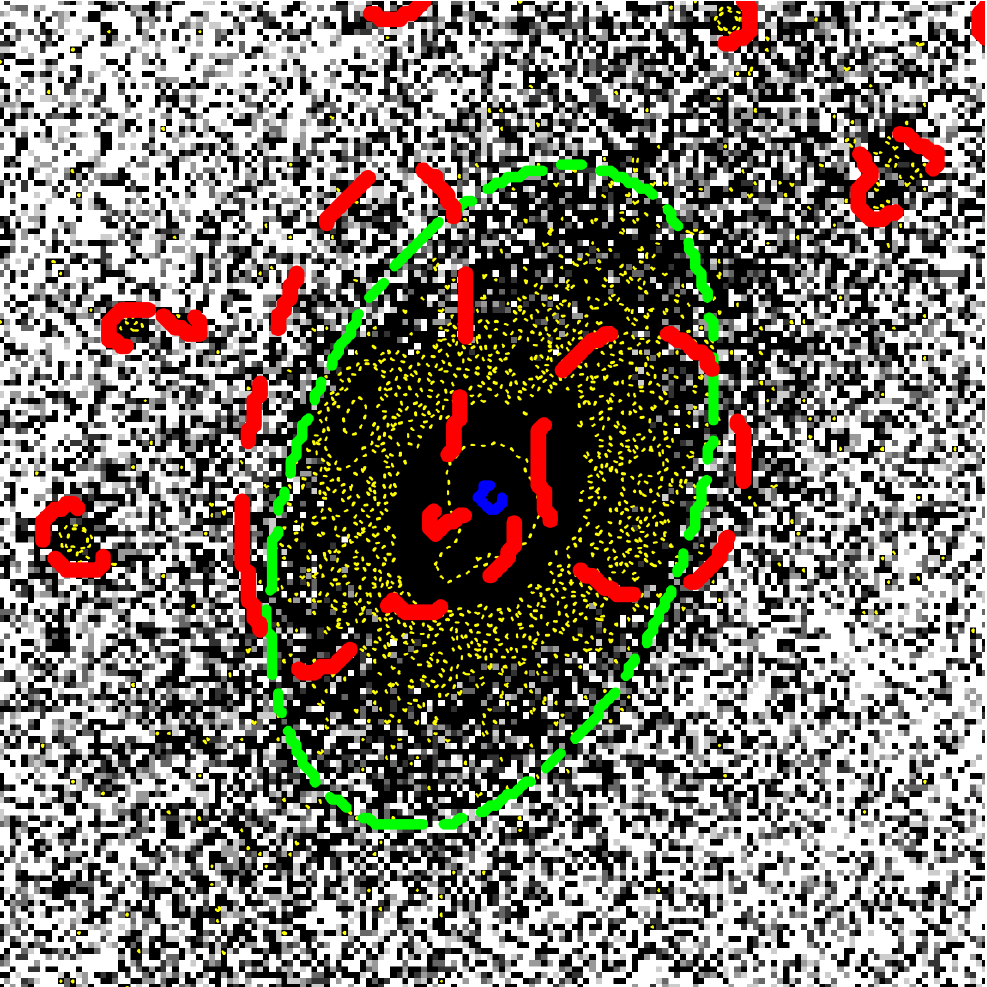}
  \plotone{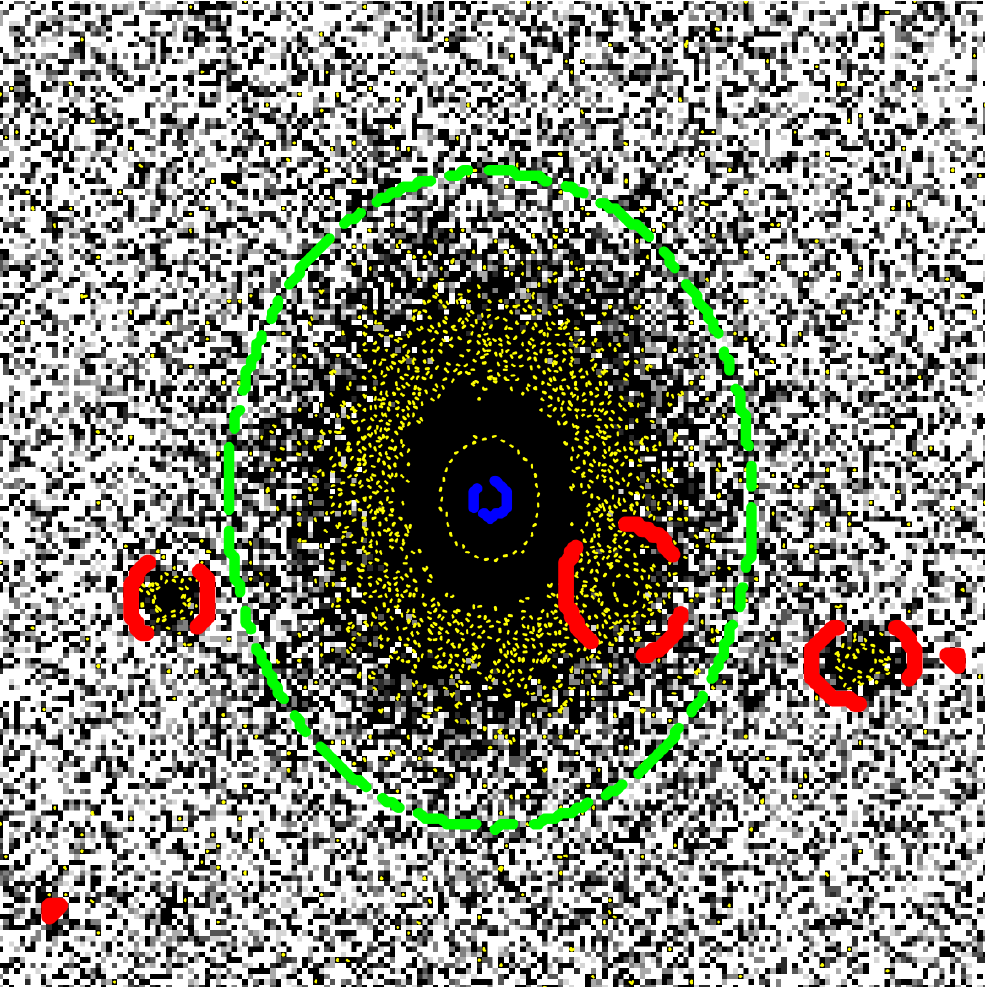}
  \plotone{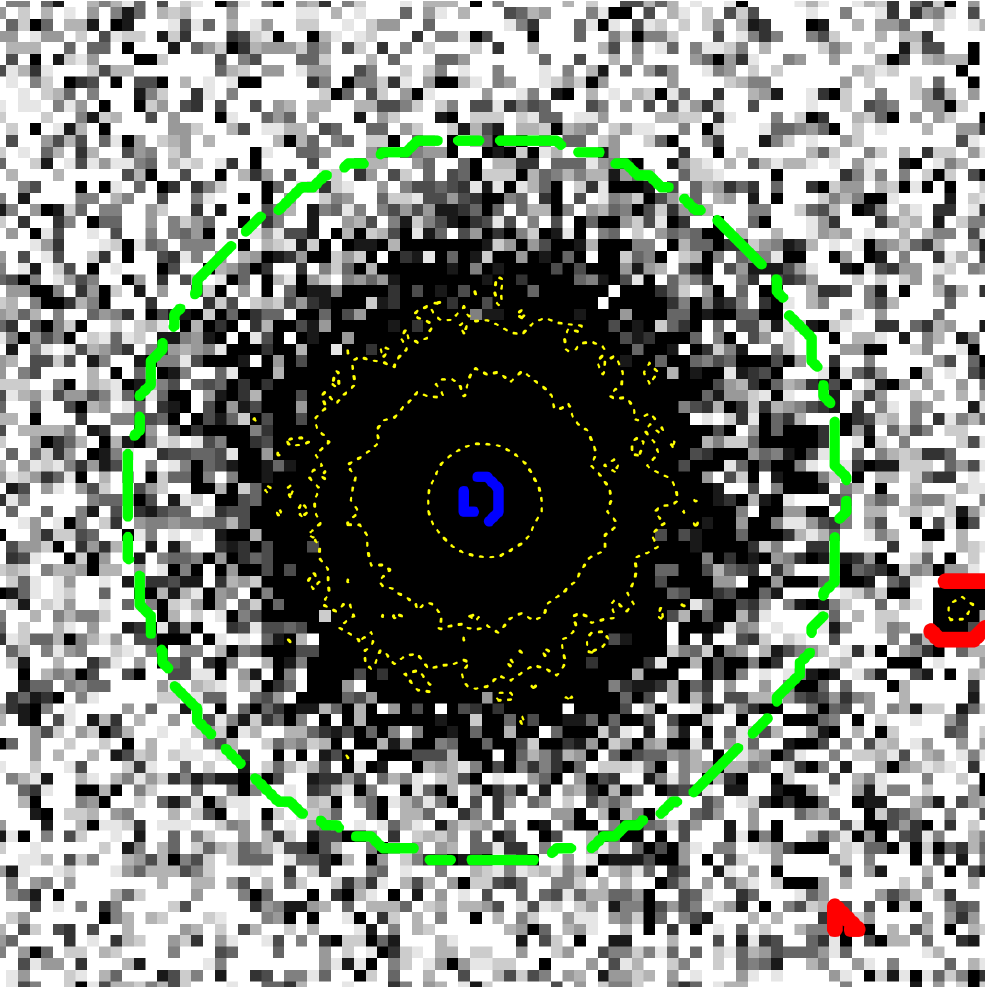}
	
  \caption{ From left to right, this figure shows examples for each
    SQ. The first image corresponds to a galaxy that is close to the
    frame border (SQ = BORDER); the second one to a galaxy 
    contaminated by secondary sources extending to the central region (SQ =
    OVERLAP\_CENTER); the third one to a galaxy where a secondary object
    is inside the green ellipse but does not overlap with the
    central region (SQ = OVERLAP\_SOURCE). Finally, the last figure shows a
    clean image (SQ = OK). Dotted red lines indicate the
    secondary source masked area. Dotted green lines indicate the
    objective galaxy.\label{stampTd}}
\end{figure}

Finally, the SQ flag is assigned based on the position of the
secondary objects relative to the target galaxy. Let $R_{i}$ be the
distance from the secondary object to the target.  We assign \(SQ=3\)
if a secondary object overlaps the central region, i.e.
\(R_{i}<\mbox{FWHM}\).  We assign \(SQ=2\) if a secondary object is in
the unmasked region with \(R_{i}>FWHM\) and \(R_{i} < \mbox{target
  major axis}\).  We assign \(SQ=1\) if we can not create a square
image stamp with 15$\times$\texttt{deVrad} on a side (see some
examples in Figure \ref{stampTd}) without intersecting the frame
border. The SQ values are summarized in Table \ref{stampoverlap}.

\begin{table}[ht]
  \begin{center}
    \caption{Stamp quality flag (SQ)\label{stampoverlap}} \vskip 0.2 truecm
    \begin{tabular}{@{} lcc @{}}    
      \hline\hline
			
      CRITERIA& NAME & Flag \\
      \tableline
      Clean stamp& OK &  0  \\ 
      Galaxy objective close to the FRAME edges& BORDER &  1  \\ 
      Secondary objects over the source& OVERLAP\_SOURCE &  2  \\ 
      Secondary objects over the central region & OVERLAP\_CENTRAL &  3 
      \\
      \tableline
    \end{tabular}
    
  \end{center}
\end{table}

\subsection{Running GALPHAT on a HPC cluster}

PyPiGALPHAT reads the input galaxy catalog and submits the jobs to the
HPC cluster. Each job is responsible for the processing of one
galaxy. By default, PyPiGALPHAT will process all galaxies in the input
file, but the user can optionally choose a single galaxy from the
list, e.g. using the SDSS \texttt{objid}.  There are three running
modes: (1) perform a new inference; (2) resume a pre-stored
inference; and (3) run the pre-postprocessing steps to verify
successful completion.  This last mode identifies the stages of the
pipeline that have failed and reruns the inference if necessary.

The posterior sample of approximately 100,000 states generated by the
MCMC inference needs to be managed efficiently. The ASCII state files
are converted to Flexible Image Transport System (FITS) in binary
table format.  The pre-postprocessing step removes unnecessary log
files for successful inferences.  When using the multiple-chain differential
evolution algorithm, some chains become stuck in regions of
anomalously low posterior probability.  These chains may be identified
and trimmed from the posterior sample using an outlier detection
scheme provided by the BIE.  PyPiGALPHAT has the option to perform
parameter estimation with GALFIT for comparison.  The PyPiGALPHAT
produces the setup, processes, and validates the results obtained by
GALFIT.

\subsection{Output catalogs and diagnostic plots}

PyPiGALPHAT provides preliminary analysis of each posterior
distribution.  Each job is responsible for the postprocessing of one
galaxy.  Each job calls a \texttt{R
  script}\footnote{https://www.r-project.org/}, which has been
developed to obtain the diagnostic figures and catalogs with inferred
values(e.g. MAP, ML, Median,Mean), pooled posteriors, and covariances.

Galaxies with a low quality flag (QF=1 or QF=3; see
Table~\ref{stamQuality}) can converge to incorrect stationary
solutions.  To detect this and other anomalous conditions, each
parameter in the posterior sample is offset and rescaled according to the
prior specification defined in Table \ref{priorsTable}. The
one-dimensional marginal distribution and the quantiles 25\% and 75\%
($Q25$ and $Q75$, respectively) are computed for all model parameters,
as well as the MAP and ML solutions. These inferred values can be used
to estimate the variance from the interquartile range ($\sigma =
0.74\,(Q75-Q25)$).

\begin{figure}[ht]
  \epsscale{0.3}
  \plotone{obs96.pdf}
  \plotone{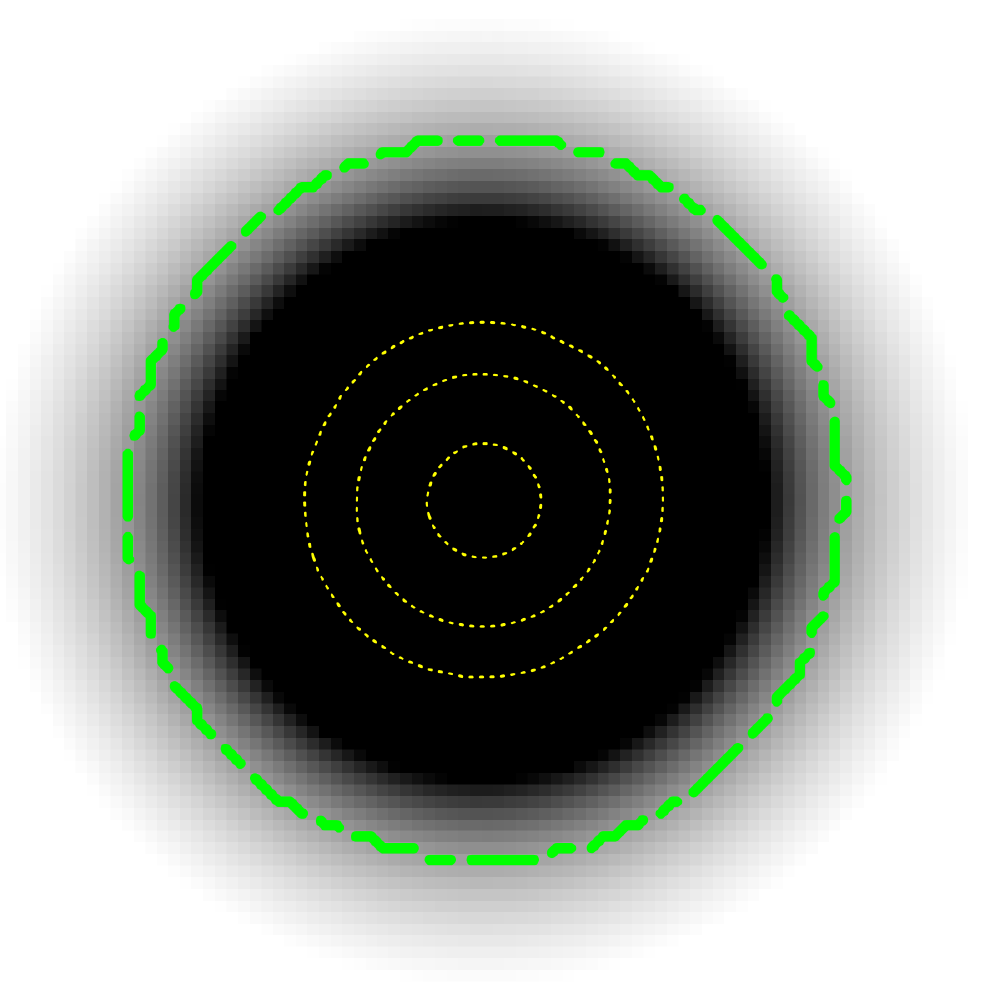}
  \plotone{res96.pdf}
  \caption{Left: an observed Stamp of a given galaxy. Green (red)
    dotted lines indicate the Petrosian region and the nearby
    secondary objects. Middle: a model image corresponding to the MAP
    solution. Right: the MAP residual that corresponds to the
    difference between the observed and model images, normalized by
    observed stamps.\label{posteriorMP}}
\end{figure}

Figure \ref{posteriorMP} shows the postage stamp, the model image, and
the residual image of a galaxy considering MAP solutions. These images
can help to rapidly identify problematic situations, e.g. incorrect
centering or orientation (position angle), mask files missing a
secondary source, etc. For each residual image we compute their
extreme values, mean and RMS values, which are then saved in the
output catalog.  All inferred quantities like quantiles, MAP and ML
solutions, covariances, likelihood marginalization, and residual
extreme values are saved in a final catalog.

\bibliography{references}

\end{document}